\documentclass[compsoc,conference,a4paper]{IEEEtran}

\usepackage{indentfirst}
\usepackage{graphicx}
\usepackage{subcaption}
\usepackage{algorithm}
\usepackage{algorithmicx}
\usepackage{algpseudocode}
\usepackage{comment}
\usepackage{enumitem}
\usepackage{color}
\setlist[itemize]{leftmargin=*}

\usepackage{hyperref}
\usepackage{booktabs}

\usepackage{amsmath, amsthm, amsfonts, graphicx, bbm}

\usepackage{soul,xspace}

\newif\ifTechnicalReport

\TechnicalReporttrue

\newif\ifCameraDraft
\CameraDraftfalse

\newcommand{\etal}{\textit{et al.}\xspace}
\newcommand{\eg}{\textit{e.g.}\xspace}
\newcommand{\ie}{\textit{i.e.}\xspace}

\newcommand{\newtext}[1]{{\color{black}{#1}}\xspace}
\newcommand{\sandb}[1]{\ifCameraDraft\color{blue}{#1}\color{black}\else#1\fi}
\newcommand{\sandbrm}[1]{\ifCameraDraft\color{red}{\st{#1}}\color{black}\fi}

\theoremstyle{definition}

\newcommand{\sm}{SpeedyMurmurs\xspace}

\newcommand{\ff}{Ford-Fulkerson\xspace}

\newcommand{\dsSF}{the scale-free dataset\xspace}

\newcommand{\src}{sender\xspace}
\newcommand{\dest}{recipient\xspace}

\newcommand{\figwidth}{0.49}

\newcommand{\expSuccPLCreditCumulative}[4]{
\begin{figure} []
  \begin{subfigure}{\figwidth\columnwidth}
  \centering
  \includegraphics[width=\columnwidth]{figures/#1/Success-Ratio-vs-Time-Epochs.png}
  \caption{\sm}
  \label{figure:#4-Success-Ratio-vs-Time-Epochs-sm}
  \end{subfigure}
  \begin{subfigure}{\figwidth\columnwidth}
  \centering
  \includegraphics[width=\columnwidth]{figures/#2/Success-Ratio-vs-Time-Epochs.png}
  \caption{\ff}
  \label{figure:#4-Success-Ratio-vs-Time-Epochs-ff}
  \end{subfigure}
  \caption{#3}
  \label{figure:#4}
  \vspace{-1.5em}
\end{figure}}

\newcommand{\plotexpsByAttack}[4]{
  \begin{figure*} []
    \begin{subfigure}{0.32\textwidth}
      \centering
      \includegraphics[width=\textwidth]{figures/#2-success_ratio.png}
      \caption{Random selection, Ford-Fulkerson}
      \label{figure:#2-success-ratio}
    \end{subfigure}
    \begin{subfigure}{0.32\textwidth}
      \centering
      \includegraphics[width=\textwidth]{figures/#1-success_ratio.png}
      \caption{Random selection, SpeedyMurmurs}
      \label{figure:#1-success-ratio}
    \end{subfigure}
    \begin{subfigure}{0.32\textwidth}
      \centering
      \includegraphics[width=\textwidth]{figures/#3-success_ratio.png}
      \caption{Tree selection, SpeedyMurmurs}
      \label{figure:#3-success-ratio}
    \end{subfigure}

    \caption{#4}
    \label{figure:#1-#2-#3}
    \vspace{-0.5em}
\end{figure*}}

\newcommand{\plotexpsCentrality}[3]{
  \begin{figure} []
    \begin{subfigure}{\figwidth\columnwidth}
      \centering
      \includegraphics[width=\columnwidth]{figures/#1-success_ratio.png}
      \caption{SpeedyMurmurs}
      \label{figure:#1-success-ratio}
    \end{subfigure}
    \begin{subfigure}{\figwidth\columnwidth}
      \centering
      \includegraphics[width=\columnwidth]{figures/#2-success_ratio.png}
      \caption{Ford-Fulkerson}
      \label{figure:#2-success-ratio}
    \end{subfigure}

    \caption{#3}
    \label{figure:#1-#2}
    \vspace{-0.5em}
\end{figure}}

\newcommand{\plotexpsComparison}[3]{
  \begin{figure} []
    \begin{subfigure}{\figwidth\columnwidth}
      \centering
      \includegraphics[width=\columnwidth]{figures/#1-success_ratio.png}
      \caption{Dropping}
      \label{figure:#1-success-ratio}
    \end{subfigure}
    \begin{subfigure}{\figwidth\columnwidth}
      \centering
      \includegraphics[width=\columnwidth]{figures/#2-success_ratio.png}
      \caption{Griefing by delaying}
      \label{figure:#2-success-ratio}
    \end{subfigure}

    \caption{#3}
    \label{figure:#1-#2}
    \vspace{-0.5em}
\end{figure}}

\begin{document}

\title{Structural Attacks on Local Routing in Payment Channel Networks}

\author{\IEEEauthorblockN{Ben Weintraub}
\IEEEauthorblockA{\textit{Northeastern University}\\
 weintraub.b@northeastern.edu}
\and
\IEEEauthorblockN{Cristina Nita-Rotaru}
\IEEEauthorblockA{\textit{Northeastern University}\\
c.nitarotaru@northeastern.edu}
\and
\IEEEauthorblockN{Stefanie Roos}
\IEEEauthorblockA{\textit{TU Delft}\\
s.roos@tudelft.nl}}

\maketitle

\begin{abstract}
Payment channel networks (PCN) enable scalable blockchain transactions without
fundamentally changing the underlying distributed ledger algorithm. However,
routing a payment via multiple channels in a PCN requires locking collateral for
potentially long periods of time. Adversaries can abuse this mechanism to
conduct denial-of-service attacks. Previous work focused on source routing,
which is unlikely to remain a viable routing approach as these networks grow.

In this work, we examine the effectiveness of attacks in PCNs that use routing
algorithms based on local knowledge, where compromised intermediate nodes can
delay or drop transactions to create denial-of-service. We focus on
SpeedyMurmurs as a representative of such protocols. We identify two attacker
node selection strategies; one based on the position in the routing tree, and
the other on betweenness centrality. Our simulation-driven study shows that
while they are both effective, the centrality-based attack approaches
near-optimal effectiveness. We also show that the attacks are ineffective in
less centralized networks and discuss incentives for the participants
in PCNs to create less centralized topologies through the payment channels they
establish among themselves.
  %% \stef{should we mention which attack is better?}
\end{abstract}

\begin{IEEEkeywords}
Cyrptocurrency; Routing Attack; Payment Channel Network; Lightning Network;
Local Routing; Centrality
\end{IEEEkeywords}

% !TEX root = creditnetworks.tex

\section{Introduction}
\label{sec:intro}

Payment channel networks such as Lightning~\cite{lightningnetwork} are the predominant solution to scaling blockchains without fundamentally changing the underlying consensus algorithm~\cite{gudgeon2019sok}. At their core, they rely on the concept of a {\em payment
channel}, wherein two parties who wish to perform transactions off-the-ledger put funds
in escrow dedicated to such operations. For parties that do not directly share a channel,
payment channel networks facilitate transactions by forwarding the payment along one~\cite{lightningnetwork}
or multiple~\cite{speedymurmurs,malavolta2019anonymous,sivaraman2020high} paths in the
graph created by all existing direct payment channels.
The chosen paths need to have the necessary funds to complete the transaction.
The process of finding suitable paths is referred to as {\em routing},
as an analogy to routing in networks, with the caveat that the main goal here is
completion of the payment.

Routing payments along one or multiple paths proceeds in two phases: First, all
involved nodes commit to paying their successor on the path. Such a commitment
implies that they \emph{lock} the funds required for the payment as
\emph{collateral}. As a consequence, these funds are not available for any
concurrent payments. Second, the payments are finalized. If the payment fails,
the collateral is released after a timeout. These timeouts tend to be on the
order of minutes or even hours, \eg, for Lightning, 40 Bitcoin blocks in the
future from the current block number (at the start of the transaction) --- more than 6 hours~\cite{ln_bolt7}.
Adversarial parties can abuse this mechanism for a denial-of-service attack. By forcing collateral to remain locked in a maximal number of channels for long periods, they can drastically reduce the funds available for other concurrent payments. As a consequence, concurrent payments can fail due to a lack of available funds~\cite{rohrer2019discharged,perez2019lockdown}.
Such an attack is called a griefing attack~\cite{ConnectorRiskMitigations}.
%% \stef{I think~\cite{interledger} is the wrong reference}

Previous work ~\cite{perez2019lockdown,rohrer2019discharged,mizrahi2020congestion}
on griefing attacks
has  focused on single-path source
routing as this is the current type of algorithm in the Lighting Network. In
such a setting the attacker is the source of the payment
and no intermediate nodes are involved in the attack. Initiating
payments for an attack is \sandbrm{both more}costly\sandb{, however}\sandbrm{ and more likely to raise suspicions
due to a high amount of (failed) payments originating from the same source}.
The attacks exploit either the specifics of Lightning’s source routing protocol
by selecting a cheap path, or leverage properties of the underlying PoW
blockchain (\eg, block size, transaction limit).
%has explored griefing attacks
%  Tochner
%  \etal~\cite{tochnerRouteHijackingDoS2020} present an attack specific
%  to Lightning’s source routing protocol; the attack
%  does not translate to local routing
%  as it leverages the fact that the source selects a cheap path. Both Mizrahi
%  \etal~\cite{mizrahi2020congestion} and Harris \etal~\cite{harris2020flood}
%  leverage properties of the underlying PoW blockchain (\eg, block size,
%  transaction limit) for their attack.

It is unlikely that source routing will remain the routing in the Lightning Network
as network usage increases. Not
only does source routing require storing a snapshot of the global topology at
each node, it also prevents intermediary nodes from adjusting the path if a
channel does not have sufficient funds, thus leading to routing failures even in
the absence of attacks~\cite{sivaraman2020high}. A na\"ive solution would be to
maintain information about current funds at the source. However, keeping such
information up-to-date is infeasible in a large network, because each successful
transaction entails changes in the available funds of one or more channels. In a
network of one million nodes, every update due to a transaction has to be
forwarded at least one million times. If PCNs, indeed, settle thousands of
transactions per second, as the VISA network
does~\cite{visa_fact_sheet},
billions of update messages have to be sent per second; an unacceptable
overhead.

%% As a consequence, source routing is not a suitable solution for large-scale payment networks.

To overcome the limitations of source routing, routing algorithms solely based on local information have been developed~\cite{silentwhispers,speedymurmurs,Zhang_Yang_2021}. Such approaches allow intermediaries to choose suitable channels based on the currently available funds. However, giving more power to intermediate nodes also opens these routing algorithms to attacks. Prior work  did not evaluate the effect of denial-of-service attacks on such routing algorithms.
Given the severity of such attacks for source routing, it is essential to evaluate novel algorithms in the presence of such attacks before deploying them. To the best of our knowledge such attacks have not been studied for payment channels routing with local knowledge.

%Finally, an attacker can take advantage of publicly available information to amplify their attacks.
%Several new proposals for routing algorithms in payment channel provide privacy of the transaction
%value. In such settings an attacker can not leverage transaction information to select high-value transactions and amplify the attacks.

In this work, we focus on attacks against local routing algorithms in payment channel networks,
where the attacker is an intermediate node on a payment path.
More precisely, we perform two versions of a denial-of-service attack on
\sm~\cite{speedymurmurs}, a routing algorithm based on local information, which
is considered \sandb{a } promising alternative to source routing~\cite{gudgeon2019sok}.
In the first variant of our attack, the attacker drops payments entirely. In
the second variant, the attacker performs griefing by delaying the payment
without causing it to fail.
%% In the second variant, the attacker drops the payment entirely.
% ben: removed sentence below, because we do not model timeouts in our drop attack
%% Due to the locked collateral, such an attack does not only affect the current payment. All funds locked by the attacker's predecessors on the paths remain unavailable until the respective timeouts expire, which can cause concurrent payments to fail.
All attacks are performed by intermediaries rather than the source, with the intermediaries being selected strategically based on their position in the network. To conduct the attacks, the attacker needs to know the topology of the network. Note that attackers cannot be prevented from learning the topology as in all currently deployed PCNs
(including \sm), channel opening and closing are recorded on the public blockchain.

In contrast to source routing, \sm's ability to let intermediaries detect and avoid channels with blocked collateral should make it more resistant to such attacks. Our simulations show the vulnerability of \sm~to attacks.
We observe, in all simulated scenarios, that dropping is more damaging than
griefing. However, network operators can more easily detect dropping than
griefing. This is because in a griefing attack, a transaction may fail due to an
attacker delaying a transaction on a partially overlapping path, whereas with
dropping, the transaction must be directly routed through the attacker's node.

%We show that
%SpeedyMurmurs' performance degrades slowly when attackers drop or grief payments.
%However, by using publicly available information about the graph connectivity as defined by the payment
%channels, the attack can be improved significantly.
%Indeed, by selecting as attackers, the 0.1\% of nodes with the highest betweenness centrality, we cause
%as many failures as if a full 30\% of nodes in the network were selected randomly.

Our results indicate that selecting attackers by graph centrality is the most
effective selection method. Specifically, a centrality-based attacker must
corrupt just 0.1\% of the nodes to reduce the fraction of successful
transactions to near zero, which is only slightly less effective than an ideal
attacker that selects the nodes based on the number of transactions they relay.
In contrast, our \sm-specific attack, which selects attackers based on spanning
tree-depth, requires 3\%, while a random node selection requires 20\% to do the
same degree of damage. Though \sm is a tree-based algorithm, it allows
the use of channels not in the tree. Thus, nodes in central positions in the
tree do not forward as many transactions as nodes with a central positions in
the graph, leading to the lower effectiveness of the tree-based attack.

We perform a cost-analysis and show that our most powerful attack,
which requires only 10 attackers in a network of 10,000,  would
cost an estimated one million USD to perform. This may seem expensive, however,
as our simulated network resembles the size of Lightning, which as
of Mar. 22, \sandb{2021}~\cite{LightningNetworkStatistics} has a capacity
of 64 million USD,
the lost income
due to nearly all the transactions
failing
could dwarf the attack cost.

Based on the observation that centrality drastically increases the impact of the attacker,
we evaluate the attacks in a small-world network with homogeneous node centralities and
find that the attack is indeed less effective.
Consequently, we discuss incentives for nodes to transform existing payment channel networks to more suitable topologies.

%% delaying alone causes as much damage all payments to fail if the 0.1\% nodes with the highest betweenness centrality are selected as attackers.
%% and dropping
%% \stef{1) claim that we only need direction is not explained}

%A side outcome of our work includes a method for creating configurable and widely applicable datasets for evaluating PCNs. The existing datasets are missing
%information~\cite{prihodko2016flare} or are based on real-world datasets from different applications such as credit networks or credit card payments~\cite{speedymurmurs,rohrer2019discharged,bagaria2019boomerang,sivaraman2020high} whose applicability is unclear.
%In contrast, our method offers a range of options for creating the underlying network graph, selecting transaction values, denoting pairs of parties performing transactions, and assigning initial balances for the channels in the graph.
%\stef{would not necessarily go into details on side contribution.}\ben{Should we
%remove this paragraph since we are not really going in-depth on this anymore?}
%In this manner, we can create datasets with varying success ratios and clearly differentiate between the effect of the attack on a network with sufficient capacity to complete all transactions and one that has strong capacity limitations.
%Generating such data sets is of use for payment channel network research as a whole.

In summary, our contributions are:
\begin{itemize}
\item A design of dropping and griefing attacks specific to local routing with
  intermediaries as attackers. 
\item A simulation-based evaluation of the proposed attacks revealing that
  high-centrality nodes allow for highly effective attacks. The strongest attacker
  uses just 0.1\% of such nodes to reduce the fraction of successful
transactions to near zero, close to what an ideal attacker could do.
\item A cost-analysis of our attacks with current transaction pricing, indicating that the attack would cost 1 million USD in a network worth 64 million USD. 
\item A strategy for reducing a network's susceptibility to our attacks by incentivizing nodes to build less centralized topologies.
\end{itemize}

% !TEX root = creditnetworks.tex

\section{Payment Channel Networks}
\label{sec:pcn}

In this section,  we introduce the key ideas of PCNs as well as components of PCNs relevant for the remainder of the paper.

\subsection{Payment Channels}
A \textit{payment channel} defines the relationship between two parties who wish to perform monetary transactions in a common digital currency. In the most general form, a channel is defined by the two parties that establish it and the amount of funds that they make available for transactions to each other.
There are three operations that can be performed on each payment channel:  i) opening the channel, ii) performing transactions, and iii) closing the channel.

\begin{figure} [!ht]
  \centering
  \includegraphics[width=0.7\columnwidth]{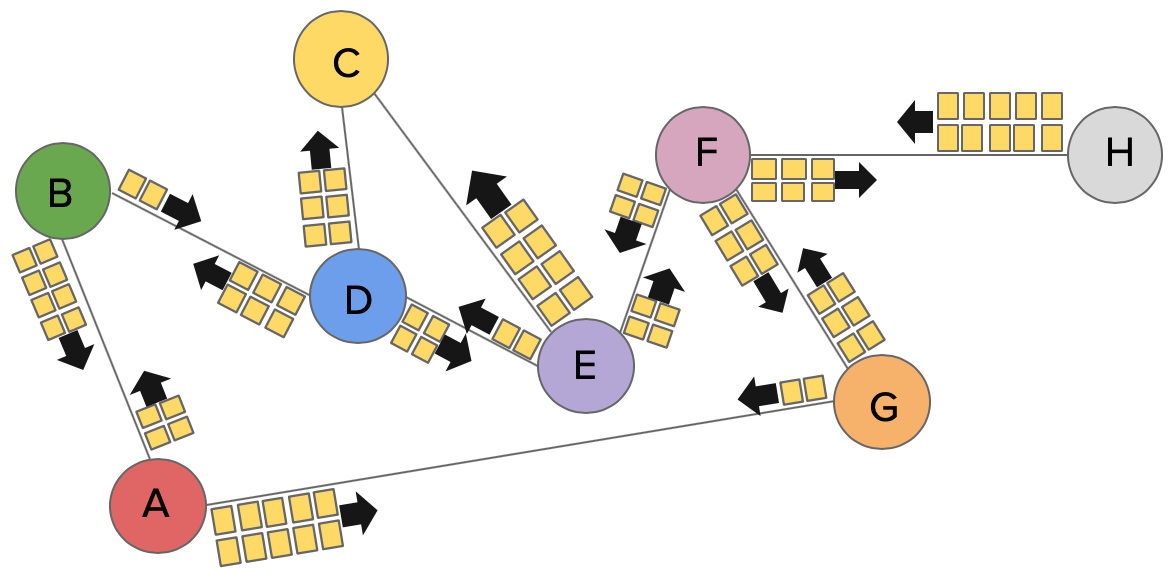}
  \caption{A payment channel network.}
  \label{fig:pcn}
  \vspace{-0.3cm}
\end{figure}

{\em Open channel.}
The \emph{channel balance} is initially established by a \textit{channel open} operation. This operation may be a verified transaction as is the case for Bitcoin's~\cite{bitcoin} Lightning Network~\cite{lightningnetwork} or Ethereum's~\cite{ethereum} Raiden Network~\cite{raidennetwork}, where channel creation is a blockchain transaction that uses smart contracts to hold the party's funds in escrow. Payment channels may be either bidirectional or unidirectional. We focus on bidirectional payments channels, \ie, payments can be sent in either direction.
A diagram of a payment channel network can be found in Figure~\ref{fig:pcn}.

{\em Perform transaction.}
A \textit{transaction} in a payment channel is initiated by one party, referred to as the \emph{\src}, proposing a new state of the channel to the other party, referred to as the \textit{\dest}. The transaction changes the \emph{balance} of the channel, \ie, the amount the \src can send to the \dest.
There are a number of mechanisms that enable secure transactions on a
channel~\cite{gudgeon2019sok} --- mechanisms that ensure the \dest receives exactly the promised funds.

{\em Close channel.} Either party on the channel can decide to close the
channel. When closing the channel, one or both parties publish the latest state
of the channel on the blockchain to regain the coins corresponding to the balance they have on their side.
Disputes between the two parties are resolved by the blockchain consensus~\cite{gudgeon2019sok}.

\subsection{Payment Channel Networks}
\label{sec:pcnetwork}
An open payment channel requires at least one party to escrow funds.
As a result, the number of channels a party is willing or able to open is
limited. Payment channel networks were proposed to facilitate payments between
parties who do not have a direct channel between them. If one considers the
parties and the channels between them as a graph, then as long as at least
one path with enough liquidity exists between the two parties, they can conduct
transactions without having a direct channel by performing a series of pairwise
transactions along each channel of each path. If a payment is split over
multiple paths, the sum of the partial payments must equal at least the total
payment value.

A brief note on semantics: we will follow the convention of
Bagaria \etal~\cite{bagaria2019boomerang} and use the term \emph{payment} to
indicate the high-level task that a user might wish to accomplish, and the term
\emph{transaction} to mean the components that make up that payment; these
components include individual hops along a single path and also payment splits
in multi-path routing.

Depending on the implementation, there are various mechanisms in place to
guarantee atomicity so that either all of the pairwise transactions succeed, or
none of them do~\cite{lightningnetwork,atomicmultipath,malavolta2019anonymous}.
These typically proceed in two rounds: During the \emph{commitment} phase, all
involved parties agree to participate in the payment using a smart contract that
enforces cooperation later on. In the \emph{payment} phase, parties then
finalize the payment if all parties agree to make the commitment. Otherwise,
the commitments expire after some time and the parties are able to use those
funds for other payments.

When making a transaction, both parties must lock the value of that transaction
as collateral which cannot be used for other, concurrent transactions. The key
idea of a griefing attack is to have parties lock collateral for longer than
intended periods of time. In this manner, the attacker prevents benign
transactions from succeeding as the locked collateral is not available and hence
the liquidity of the network reduced~\cite{rohrer2019discharged}.

\subsection{Routing Algorithms}
\label{sec:routingalgo}
Finding payment paths is one of the core challenges of PCNs. Several algorithms
have been proposed with different properties and
goals~\cite{lightningnetwork,prihodko2016flare,silentwhispers,
  speedymurmurs,sivaraman2020high,dong2018celer}. Many of the algorithms use
source
routing~\cite{lightningnetwork,prihodko2016flare,silentwhispers,
  speedymurmurs,sivaraman2020high,dong2018celer}. Of the remaining algorithms,
Flare~\cite{prihodko2016flare} seems unable to deal with network dynamics and
Celer~\cite{dong2018celer} has not been evaluated for more than 100 nodes. The
only algorithms based on local information with an in-depth analysis are
SilentWhispers~\cite{silentwhispers} and \sm~\cite{speedymurmurs}. Both
algorithms provide various privacy properties with \sm showing considerably
better performance~\cite{speedymurmurs}.

Hence, we choose \sm for our attack investigation and use \ff as a baseline. \ff
makes a good baseline for the success ratio, however, it results in an
unacceptably high overhead to be a suitable algorithm in
practice~\cite{speedymurmurs}. In the following, we describe \sm in detail, for
more information on an implementation of \ff suitable for PCNs, please refer
to~\cite{speedymurmurs}, which is also the implementation we use for our
evaluation.

{\bf \sm.}
\sm~\cite{speedymurmurs} is a privacy-preserving
routing algorithm for PCNs based on local knowledge; it consists of three stages. In the first
stage, $n$ spanning trees are created.
The number of spanning trees corresponds to the number of paths a payment can use.
Increasing the number of spanning trees may improve the success ratio
and privacy properties but comes at a cost of performance as overhead operations
will increase as well. \sandb{The {\sm} protocol uses Perlmans's distributed algorithm
  for building spanning trees~\cite{perlman1985algorithm}. The protocol starts by selecting the root nodes
  the details of which are not included in the {\sm} paper, but are covered in
  works such as Byrenheid~\etal~\cite{byrenheid2020attack}. Next, the nodes in
  the network organize into a spanning tree with each node connecting to another
  that is already in the spanning tree --- searching through the topology until
  it finds such a node. The newly joined node then alerts its neighbors of its
  connection and the index of the tree it is connected to.}

In the second stage, which can be interleaved with the spanning tree
generations, nodes construct a network embedding for each spanning tree, \ie,
each node receives a coordinate\sandb{ from its parent }for each spanning tree
based on its position in the respective tree.
These coordinates enable defining a distance between two nodes, $U$ and $V$, that corresponds to the length of the path when restricted to the spanning tree.
Concretely, for each tree, the root node has the empty vector as coordinate. A child adds a random 64-bit number to the vector representing its parent's coordinate to form its own coordinate. The shortest path between two nodes in a rooted spanning tree is the sum of the length of the paths to their least common ancestor in the tree.
Let $|u|$ denote the length of a coordinate $u$ and $cpl(u,v)$ be the common prefix length of coordinates $u$ and $v$, \ie, the number of leading elements they have in common. Then the shortest path length in the tree is a distance function $d$ with
\begin{equation}
d(u,v) = |u| + |v|-2cpl(u,v).
\end{equation}

In the third stage, transactions are routed through the network. The routing
algorithm first splits the payment into $n$ randomly sized shares and then
routes each of them along a different spanning tree. Nodes forward the
transaction shares to whichever neighbor is closest to the \dest according to
the coordinate distance of the respective spanning tree, also taking care the
channels used have sufficient liquidity. Note that this path-finding
algorithm need not follow \emph{only} spanning tree channels; a node
should choose its direct neighbor that is closest to the \dest node, which might
not be a parent or child in the spanning tree. Channel balances are decreased by
the value of the payment that is routed through them. If the balance of a channel
reaches zero, then the channel is
removed from all spanning trees. \sandb{A node that is connected to the spanning
tree by a now zeroed channel will leave the spanning tree and reconnect with a
non-zeroed channel. }The affected subtrees adapt locally by
choosing alternative channels; this process is called rebalancing.

As stated above, local routing algorithms are needed to increase the scalability
of payment channel networks. However, none of the existing local algorithms has
been evaluated in terms of providing availability in the presence of adversarial
nodes. Maintaining a high degree of availability in the presence of attacks is
of the utmost importance, hence our work focuses on providing the necessary
analysis and experimental evaluation of local routing algorithms in the presence
of attacks.

% !TEX root = creditnetworks.tex

\section{Attacks against PCNs}
\label{sec:attacks}

In this section, we describe our threat model and attacks we consider in this work.
We focus on {\em internal}, malicious attackers that aim to undermine the availability of the payment service. In other words, the attacker wants to maximize the fraction of failed payments.
Adversaries aiming to abuse the protocol for monetary gain or to learn
confidential information have been addressed in previous
work~\cite{ersoy2020profit,speedymurmurs,malavolta2019anonymous}.

\subsection{Threat Model}
\label{sec:threat}
%% \cnr{here list assumptions about the attacker, resources, what it controls,
%% can it coordinate, etc; moved here text from methodology}
{\bf Computation capabilities.}
We consider an internal, active, colluding attacker that is computationally bounded.
More precisely, the attacker has powerful computational resources but cannot break cryptographic primitives.

\newtext{{\bf Incentives.}
  We assume that malicious actors may be interested in damaging the system without
  monetary gain. This could include nation-state actors attempting to destabilize
  an economy.}

{\bf Collusion.}
An attacker can create nodes that they fully
control, and they are able to corrupt formerly honest nodes through
means such as social engineering. Attacker nodes are geographically distributed in
arbitrary locations and collude with each other. Colluding nodes communicate out-of-band, which might be faster than PCN information. Thus, we assume that an attacker node is aware of any information gathered by other adversarial parties.
As motivated in previous work, nodes that are created by the attacker may be
arbitrarily connected to other nodes in the network, even those controlled by
honest participants~\cite{avarikioti2019ride,ersoy2020profit}.
However, they have no access to an honest node's locally stored information.

{\bf Attacker knowledge (topology, initial channel balances).} As is common in
PCNs, an attacker is aware of the complete network topology \sandb{--- all of
  which is openly readable on the blockchain --- this }\sandbrm{, which}includes all
connections and their \emph{initial} capacities, but not any transactions or
capacities changed by transactions. While the attacker is aware of the initial
\sandbrm{balances}\sandb{capacities}, it may not have up-to-date information about the \emph{available}
balances on channels to which it is not directly connected. This is because
while the initial balance of a channel is public information that is published
on channel opening, transactions between two connected parties can
change the channel balance without publishing updates (unless a dispute or
channel closure occurs). Thus, a node can never be sure of the
balance of a channel it is not part of. Rather, it can only say that it is
between zero and the total capacity of the channel.

If SpeedyMurmurs is used, nodes are assigned coordinates based on their position in spanning trees. When the attacker establishes a connection, it learns the coordinates of its neighbors for all trees.
As every node adds one element to its parent's coordinate, the coordinate length corresponds to the level of the node in the tree. Thus, the attacker can know how close it is to the root.

The attacker does not know the complete tree and cannot necessarily map nodes to
coordinates. If the attacker does not have a connection to a node, it cannot tell which of the neighbors of the node are its parent (unless it has a connection to all but one).

Furthermore, the attacker knows the routing algorithm and its properties. For
instance, prior work showed that nodes close to the spanning tree root forward more traffic. However, the load on the root node itself is not necessarily high. Recall from Section~\ref{sec:routingalgo} that nodes forward the transaction to the node closest to the \dest in terms of coordinates.
If the \dest shares a subtree with the source, the chosen path does not contain the root node.
If the \dest and the \src are not in the same subtree, the chosen path might still not contain the root node as there might be a shortcut, \ie, a channel between the two subtrees that is not part of the tree. The probability of finding such a shortcut is reasonably high in a densely connected graph~\cite{voute}.

{\bf Attacker placement.}
We focus on \emph{on-path} adversaries that manipulate payments they are involved in.
In contrast, \emph{off-path} attackers aim to affect transactions that are not on their channels, \eg, by
 sending their own payments along certain paths or crafting the payments in particular ways.
 Off-path attackers have been discussed in detail in prior work~\cite{mizrahi2020congestion}.

%\vspace{-10pt}
\subsection{Attack Design}
Utilizing the above capabilities, the attacker has two options to explore for attack: the selection of malicious nodes in the network topology and the actions performed by these nodes.

\subsubsection{Attacker Selection}

 For the considered \emph{on-path} attackers, selecting the position of the node in the topology is closely related to the number of transactions routed by the attacker and hence the strength of the attack. An obvious strategy is
 random selection (referred to as a {\em Random Attacker}). Several other selection strategies are particularly interesting:

 {\bf Graph-oriented.} There are many ways to quantify a node's
position within a graph \eg, connectivity,
centrality~\cite{freeman1977set}, communicability~\cite{communicability},
etc. An attacker can choose to optimize for one or several of
these properties, and choose their location in the graph accordingly.

{\em Centrality-based Attacker.}
Of particular interest to us is
\emph{betweenness centrality}~\cite{freeman1977set}, which for a specific node, $z$,
is the ratio of shortest paths between every pair of nodes that include $z$.
The betweenness centrality $c_b(z)$ of node, $z$, is given by

\[
  c_b(z) = \sum_{s,r,z \in N}^{} \frac{\sigma_{srz}}{\sigma_{sr}}
\]

where $\sigma_{sr}$ is the total number of shortest paths between
$s$ and $r$, and $\sigma_{srz}$ is the number of shortest paths between
$s$ and $r$ that include $z$.
In a routing protocol that always selects the shortest path, the betweenness centrality hence correlates with the fraction of payments forwarded via a node $z$.

{\bf Routing algorithm-oriented.}
Routing protocols typically do not select the shortest path due to constraints
such as lack of liquidity or knowledge of topological information. As such, it makes sense to adapt the attack to the path selection
of the routing protocol. Concretely, the path selection of the routing algorithm may prefer certain nodes, which should then be chosen for corruption.

{\em Tree-based Attacker.}
We can exploit that \sm is based on spanning trees~\cite{speedymurmurs}.
As stated in Section~\ref{sec:threat}, corrupting the root node is not necessarily the optimal strategy in terms of transactions an on-path attacker can affect. Furthermore, there exist protections against attacks on root nodes~\cite{byrenheid2020attack}.
Instead, the attacker can easily obtain a position close to the root. Initially, the attacker connects to random nodes. If it does not obtain a position close to the root, it connects to all neighbors of its parent.
These include the parent's parent, which will elevate the attacker by one level. The process is then continued until the attacker is connected to the root. Once the attacker knows the root, it can establish a connection to it with multiple nodes.

{\bf Transaction-oriented.}
We also considered selecting nodes based on the balance of their channels and the rate of changes. However, such information is only explicitly known to the parties in the channel.
Thus, the information should only be available to the attacker after corrupting the node and it does not seem sensible to base the selection strategy on such unknown information in a real attack.

 {\em Ideal Attacker.}
Given that corrupting the nodes based on the cumulative value of transactions handled is the strongest attack, we consider it as an {\em Ideal Attacker} and use it as baseline for comparison in our evaluation.

\subsubsection{(On-path) Attacker Actions}
\label{attacker:actions}
We consider insider attacks, and are primarily interested in intermediary nodes
on the path as they are less detectable. \newtext{For a node to have a valid
  suspicion of an attacker, it is enough to simply know which direction the
  failed payments came from --- not necessarily the source.}

There are various actions an attacker could take when it is on-path.
This is in contrast to Rohrer \etal~\cite{rohrer2019discharged} who
fully remove their selected nodes from the topology, whereas
our selected nodes perform malicious actions, which are harder
to detect.

{\bf Dropping.} An attacker could simply drop a transaction that
it \emph{should} forward to the next node in a multi-hop payment.
More maliciously, the attacker could refuse to forward the
transaction, but still send a confirmation that the transaction
has been forwarded to the previous node on the path. If a
transaction is dropped during the \emph{commitment} phase
(Section~\ref{sec:pcnetwork}), all previous hops would need to
maintain their locked collateral until the transaction expires.
We follow Miller \etal~\cite{miller2019sprites}
in assuming that a rational investor's preference is to
obtain and use money now rather than later, and therefore a
forced restriction on using one's own money constitutes an attack.
Locked collateral also cannot be used to route
other transactions that may have a higher chance at succeeding, and
thereby deprives the collateral holder of potential revenue
from fees.
Note that dropping is not performed during the
\emph{payment} phase (Section~\ref{sec:pcnetwork}) by a
rational attacker, because then they would be forced to pay their
guaranteed funds to the node one hop closer to the \dest but, in
dropping, would not be reimbursed from the node one hop closer to the \src.

{\bf Griefing by delaying.} A node can wait to forward a payment to
the next node until some specified amount of time has passed or
some condition is met. For example, one such condition could be
to wait until just before the transaction times out before
forwarding it. Doing so will allow the payment to complete, but
would force the collateral to be locked up for the maximum amount
of time, which constitutes an attack.
Performing this attack inconsistently, \eg, on only 50\% of transactions,
would also make it difficult to detect because of its similarity to network
delays. \newtext{In addition, failures from this attack can be indirect and
  thus deniable by the attacker.}
This has a similar effect to \emph{payment griefing}~\cite{ConnectorRiskMitigations}, in which
an attacker sends payments to colluding nodes that then delay and drop the transaction.
Our variant is more general, however, because \emph{any} node on the transaction path can grief
by delaying, whereas in traditional griefing, a transaction has to be addressed to the
griefing node.

%In addition to those two actions, an attacker might incorrectly forward the transaction, \eg, by choosing a random node instead of one leading towards the destination. As the subsequent nodes can usually detect such an attack, \eg, by checking if they are indeed closer to the destination than their predecessor. As nodes can then discard the transaction, the attack is essentially the same as dropping and hence already treated.

%We assume the attacker will apply the actions to all transactions. In the absence of detection schemes, there is no need for more stealthy attacks, such as targeting high-value transactions.

% !TEX root = creditnetworks.tex

\section{Attack Implementation}
\label{sec:meth}

In this section, we describe the attacks we implemented. First we describe the changes
we made to the simulator and then the realization of the attacks.

%We evaluate the \ff and \sm algorithms, with
%two types of collateral, Total Collateral and Strict Collateral, and two types of attacks, griefing by delaying and dropping. The adversary nodes were either chosen randomly or based on their centrality in the network.

%In the following, we describe our simulator and detail all of the attack design decision.
\subsection{Concurrent Transactions}

We extend the sequential transaction simulator from Roos \etal~\cite{speedymurmurs} to build
a \emph{concurrent} transaction simulator\sandb{, which we make publicly available~\footnote{\url{https://github.com/iowaguy/pcn-simulator}}}.  In the simulator, each channel is capable of locking collateral
for a configurable number of concurrent transactions.
This is done through the use of threading.
Each transaction is executed in a distinct thread that operates on a shared state in memory.
Transactions arrive uniformly and begin executing as soon as a thread is available.
%Each individual thread executes
For each transaction, the configured routing algorithm is executed in two steps.
First, a path is found using the methods of the configured routing algorithm.
This path discovery is done hop by hop, and with each hop, collateral along the
corresponding channel is reserved for the current transaction, which cannot be
violated by any other concurrent transaction. To simulate the effects of network
delays, a \emph{simulated network delay} of 30 milliseconds is used for each hop
during path finding. The delay of 30 ms was chosen as it is similar to delay
within Europe or the US (delays vary between 30 to 50 ms depending on the
provider). The end-to-end delay is bigger than 30 ms as it is multi-hop. We
experimented with a variety of delays, but the results were similar since the
dominant factor is the griefing delay of 10 seconds.

Second, when the routing algorithm has found a path with sufficiently high
balances and has locked collateral on each channel, it can then complete the
transaction in reverse order. This payment phase involves unlocking the
collateral and updating the channels' balances.

%The challenge in this comes from the frequent, and concurrent, updates to the shared channel balance state.
Not only do channel updates occur on every completed transaction, but they also occur during
collateral locking.
It is necessary to track how much collateral is locked, as locked collateral is not available for other transactions.
Thus, the important aspect in deciding whether routing via a channel is possible is not the balance but the difference between the balance and the locked collateral.
Locked collateral leads to paths being diverted by insufficient available funds or even to failures if a path with sufficiently high available balances cannot be found. A secondary reason for this information to be maintained
is that we need to be able to rollback transactions that have claimed some collateral but were unable to be
completed for some reason. Upon aborting such a partial transaction, the channel must be returned to its prior
state, while still retaining any changes from any other partial or complete transaction that may have
occurred concurrently.

\subsection{Attack Implementation}
We implement two types of attacks: \emph{dropping} and
\emph{griefing by delaying}, as introduced in Section ~\ref{sec:attacks}. For a dropping attack,
transactions are executed normally unless they encounter an adversarial party on
the path. If an adversary is selected as the next hop, the payment is
immediately marked as \emph{failed} and any collateral locked for that
payment will be rolled back.

For a griefing attack, upon encountering an adversary, the transaction is
delayed for a configurable amount of time. For our simulations, we find that the
ratio between \emph{attack delay} and \emph{simulated network delay} is
decisive, and therefore choose a value of attack-delay to be 10 seconds. With a
simulated network delay of 30 ms, the attack does not increase in
effectiveness with longer than 10 second delays. After the delay time has
elapsed, the payment is allowed to continue.

For each of these attacks, we
consider three attacker placement methods: \emph{random},
\emph{tree-based}, and
\emph{centrality-based}, \ie, by the fraction of shortest paths going
through a node~\cite{freeman1977set}. The networkx~\cite{networkx} Python
library was used for calculating betweenness centrality. Finally, we also
evaluate an {\em ideal} attacker placement as an oracle
for the maximum damage an attack can create on a given
dataset.

%\latest{When we do \emph{transaction-oriented} attacker selection, we need to
%  know how many transactions each node is involved in. This is not possible to
%  know ahead of time, so we run the simulation twice. In the first run, which
%  has no attack, we count the number of transactions that each node is involved
%  in. Then we select the $n$ highest nodes by number of transactions to be the
%  attackers for the second simulation. In this case, $n$ is the number of
%  attackers we are simulating.}
%  \stef{so that is selected from a non-attack simulation with the same parameters but different randomness or the same seeds?}

We implemented the \textit{tree-based} attacker placement as follows.
As described in Section~\ref{sec:attacks}, an attacker can gain a high position in the tree
by opening connections and closing them again if he is not sufficiently close to the root.
In our experiments, we assume that the node has already achieved a favorable position by
the time the simulation starts.
Concretely, if we have $x$ attackers and $t$ trees for our attack, we select the $x$ attackers as the $x/t$ nodes closest to the root for each tree.

% !TEX root = creditnetworks.tex

\section{Evaluation}
\label{sec:eval}

In our evaluation we set out to answer three questions:
\begin{enumerate}
\item \label{q:1} How  effective are  dropping and griefing by delaying attacks?
\item \label{q:2} How does the attacker placement in the topology influence its attack power?
\item \label{q:3} How effective are the tree-based and centrality-based attack placement strategies when compared to the ideal attacker that has complete information about transaction values and available balances?
\end{enumerate}

%% scripts/plot_exps --experiment 57 --dataset 25 --algorithm speedymurmurs --attack_type dropping --legend '0% Attackers' '1 Attacker' '0.1% Attackers' '0.5% Attackers' '1% Attackers' --write
\plotexpsByAttack{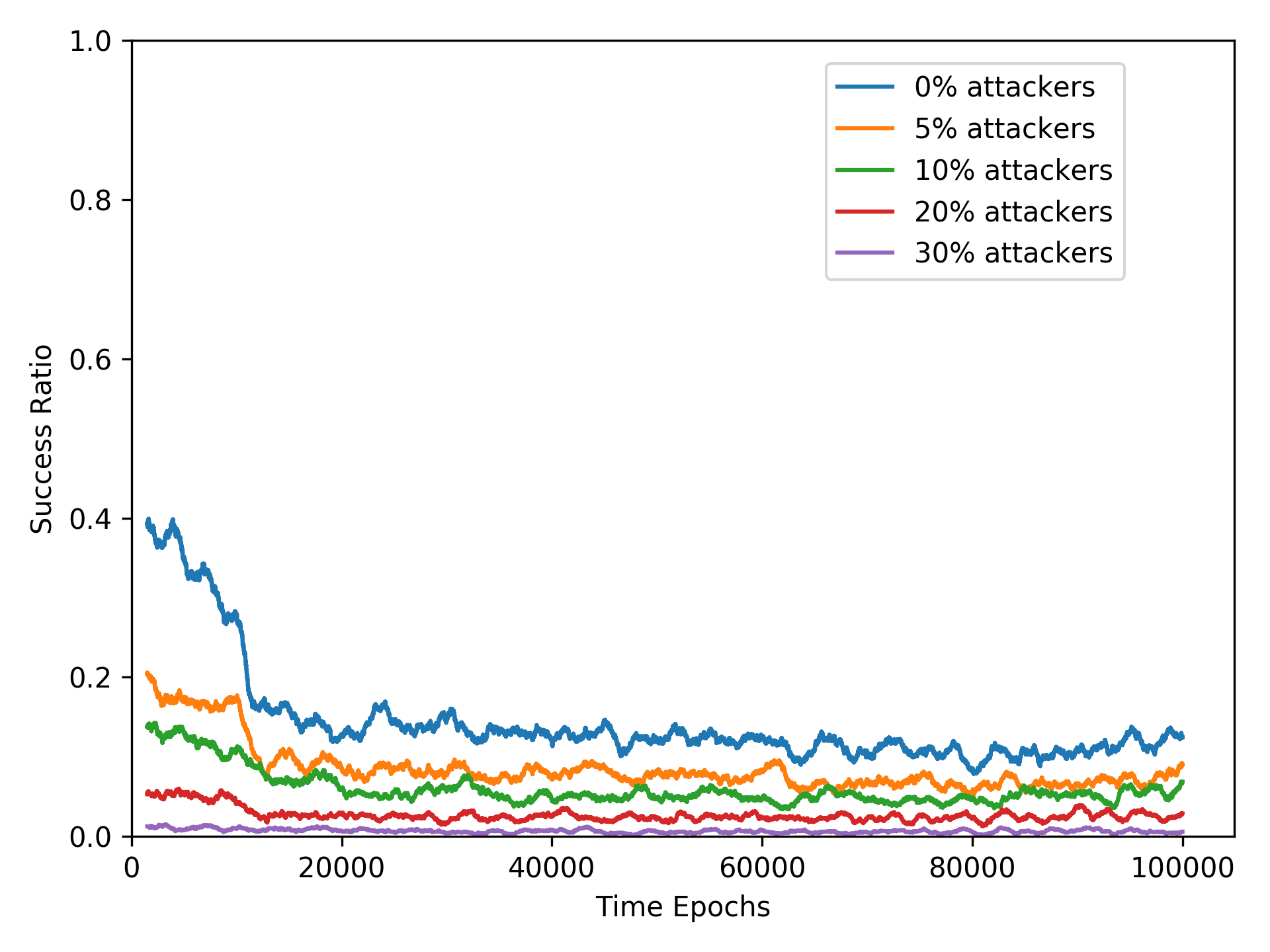}{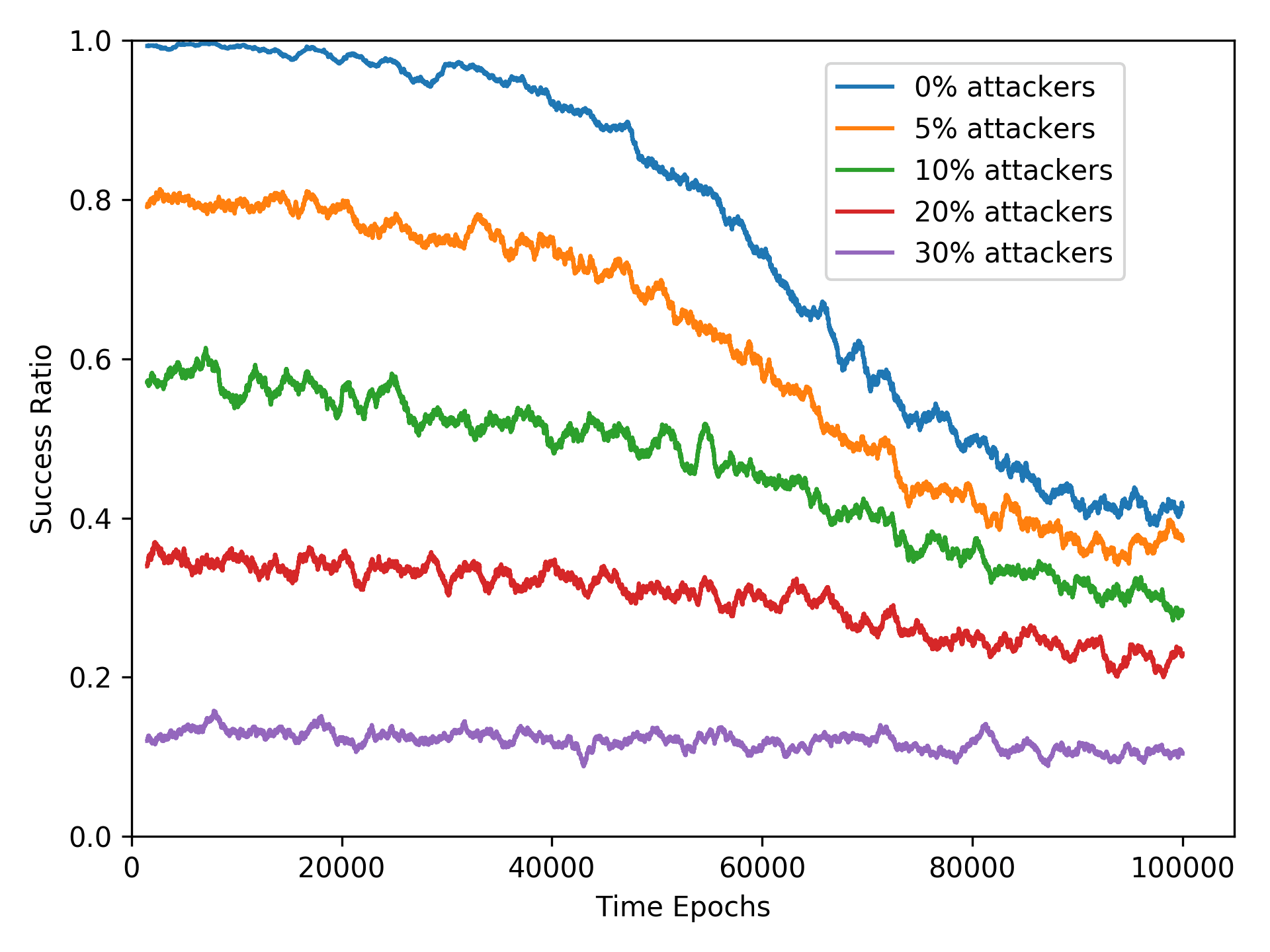}{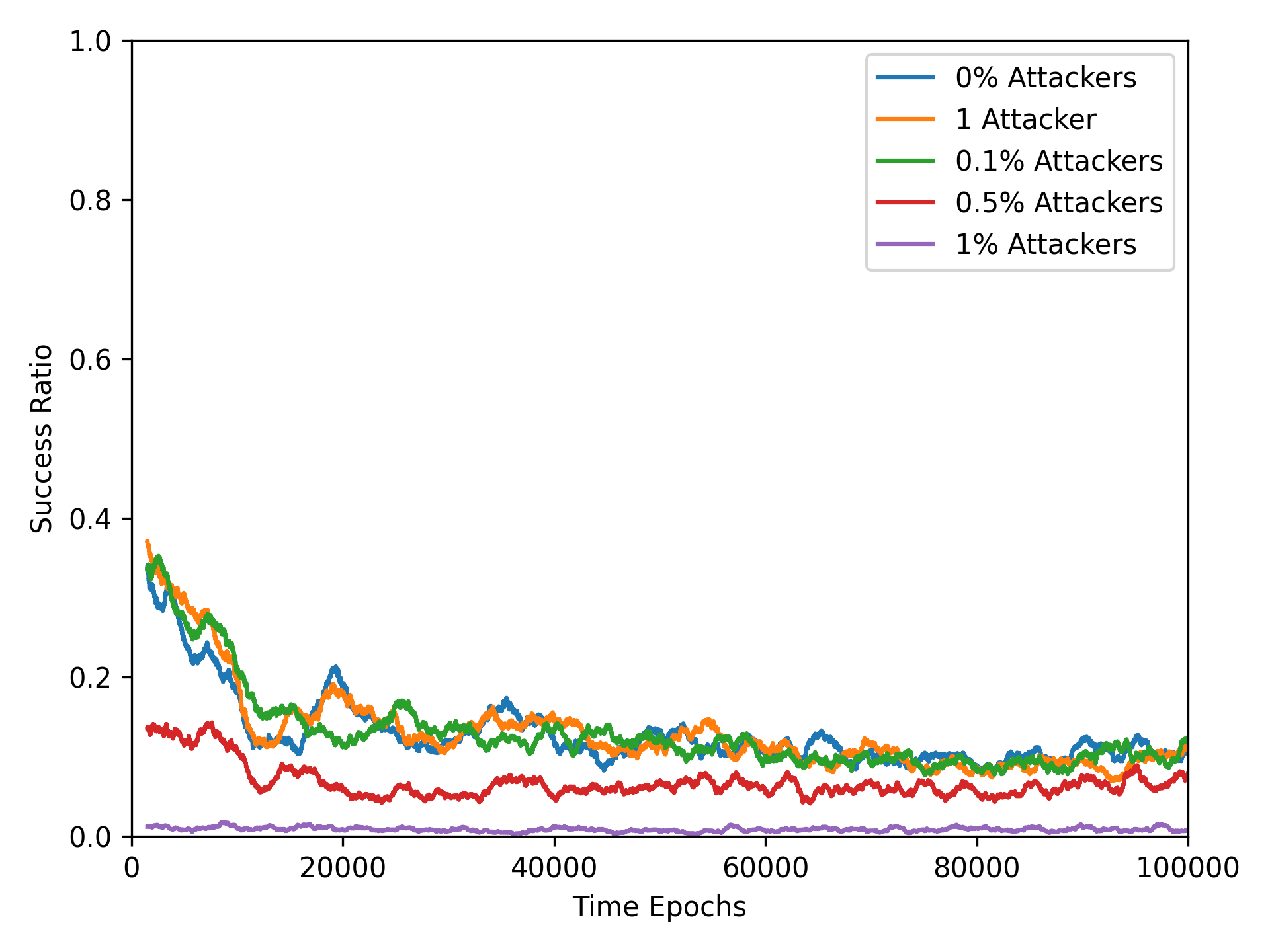}{Success
  ratios during a dropping attack.}

\subsection{Methodology}
As we are interested in the effects of these attacks on large-scale networks, we
are forced to simulate --- it would be unethical to perform such attacks on
operational financial networks. Testbeds are also insufficient for our purposes
because they do not let us evaluate at scale. In response to these shortcomings,
we generated synthetic datasets with representative network conditions to
evaluate our attacks as described in Sec.~\ref{sec:gendata}.

In all scenarios, we vary the number of attackers. For randomly selected
attackers, that amount is varied between 5--30\% of total nodes in the network.
For the centrality-based and tree-based attackers, we vary the fraction of
attackers between 0.1-5\% of nodes in the network.

We consider the success ratio to gauge the effectiveness of our attacks, \ie,
the fraction of payments for which paths with sufficient funds have been
discovered. Results for each transaction are grouped into simulated time buckets
called \emph{epochs}; this is based on when the transaction was started, not
when it finished. In other words, the results for epoch $i$ contain all
transactions initiated in epoch $i$, regardless of when they terminate. For
maximum time-step granularity, we consider each transaction to be started in a
distinct epoch unless stated otherwise. All results are presented as a running
average over 1500 epochs, unless stated otherwise. Note that our success ratios
are considerably lower than those reported in~\cite{speedymurmurs} as their data
was cleaned such that all transactions were possible. In addition, all the
presented success ratios drop over time --- this is expected because channels
become depleted as transactions traverse them. This has been previously shown
in~\cite{sivaraman2020high, khalil2017revive, linFSTRFundsSkewness2020}.

\subsection{Generated Datasets}\label{sec:gendata}
Existing datasets suffer from either a dearth of successful transactions, a lack
of information about transactions, or a small number of nodes. The Raiden
Network, for instance, has 37 nodes~\cite{RaidenExplorer} and Lighting 2337
\cite{dischargeddata}. Thus, we devise a methodology to generate representative
datasets. Our datasets consist of three components: i) the PCN topology, ii) the
transaction set, and iii) the initial channel balances.

To generate topologies, we used the ready-made implementations of the
Barab\'{a}si-Albert algorithm~\cite{barabasi1999emergence} for the scale-free
graph generation and the Newman-Watts-Strogatz
algorithm~\cite{newman1999renormalization} to generate small-world graphs. Both
implementations are from the Python module networkx~\cite{networkx}. We also
used a measured topology from the Lightning Network as collected by Rohrer
\etal~\cite{dischargeddata}.

Transaction sets were generated by sampling from representative distributions.
For the transaction values, we use a Pareto distribution as it has been shown to
reliably represent consumer spending
patterns~\cite{sandersPARETOPRINCIPLEITS1987}. Sender/recipient pairs were
generated by sampling a Poisson distribution independently for both the \src and
\dest. A Poisson distribution was chosen as prior work has shown it to be good
for modeling consumer transaction rates~\cite{faderCountingYourCustomers2005}.

Initial channel balances were generated with the following procedure. Start by
assuming that all channels have balance of zero, then calculate the shortest
path between each transaction's \src and \dest. For each channel used in the
shortest path, add the value of that transaction to its current balance. This
process is akin to starting with a fully depleted network, and then routing
transactions in reverse time. As a real network is unlikely to have the exact
amount of funds needed to route all transaction, we reduce the balances of a
fraction of channels by a 0.5 multiplier.

Unless otherwise stated, we used the \dsSF for all experiments. All datasets
used are presented in Table~\ref{tab:datasets}. More details on limitations of
existing datasets and on our generation techniques can be found in
Appendix~\ref{sec:datasets}.

\begin{table}[]
  \scalebox{0.7}{
    \centering
    \tabcolsep=0.015\linewidth
        \begin{tabular}{@{}l|l|l|l@{}}
      \toprule
      Name              & Graph Generation              &    Transaction Set                             \\ \midrule
      Scale-Free        & Barab\'{a}si-Albert           &    10k nodes, 100k transactions, 0.5 multiplier\\
      Small-World       & Newman-Watts-Strogatz         &    10k nodes, 100k transactions, 0.5 multiplier\\
      Lightning Network & Measured\cite{dischargeddata} &    2337 nodes, 100k transactions\\
      \bottomrule
    \end{tabular}
  }
\caption{Description of datasets used in experiments. Unless otherwise stated, our simulations use \dsSF.}
\label{tab:datasets}
\end{table}

\subsection{Impact of Dropping and Griefing}

We start by analyzing the dropping attack on both \ff and \sm comparing random and tree-based attacker placement.
In Fig.~\ref{figure:id25-exp22-maxflow-drop_all-success-ratio}, where \ff is
evaluated with a random attacker selection, we see significantly degraded
performance for each increase in the number of attackers with no diminishing
returns. With more attackers, the success ratio may eventually drop to zero.
As displayed in
Fig.~\ref{figure:id25-exp22-speedymurmurs-drop_all-success-ratio}, which
evaluates random attacker selection against \sm, the success ratio
is impacted significantly in the early epochs with more than 5\% attackers.
Increasing the number of attackers gives rise to a concomitant increase in the
attack's effectiveness. \sandb{Note that in both these plots, as well as all
  that follow, the success ratio has a consistently negative slope. This is
  expected because payment {\src}s and {\dest}s are selected randomly from
  non-uniform distributions, so most channels will be used more in one direction
  than the other and will, therefore, end up depleted.}

%% scripts/plot_exps --experiment 51 --dataset 25 --algorithm speedymurmurs --attack_type griefing --legend '0% Attackers' '0.1% Attackers' '0.5% Attackers' '1% Attackers' '3% Attackers' --write
\plotexpsByAttack{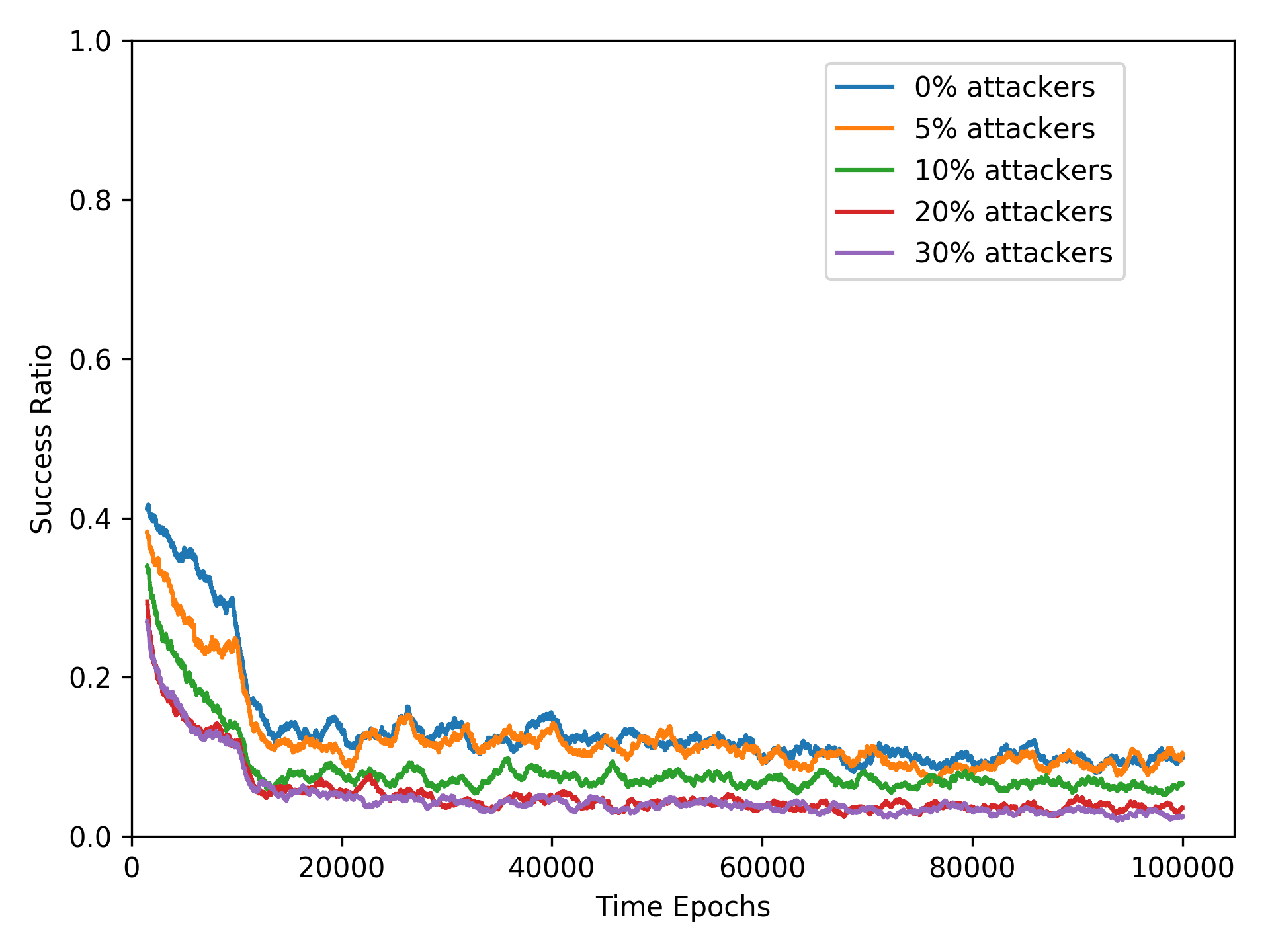}{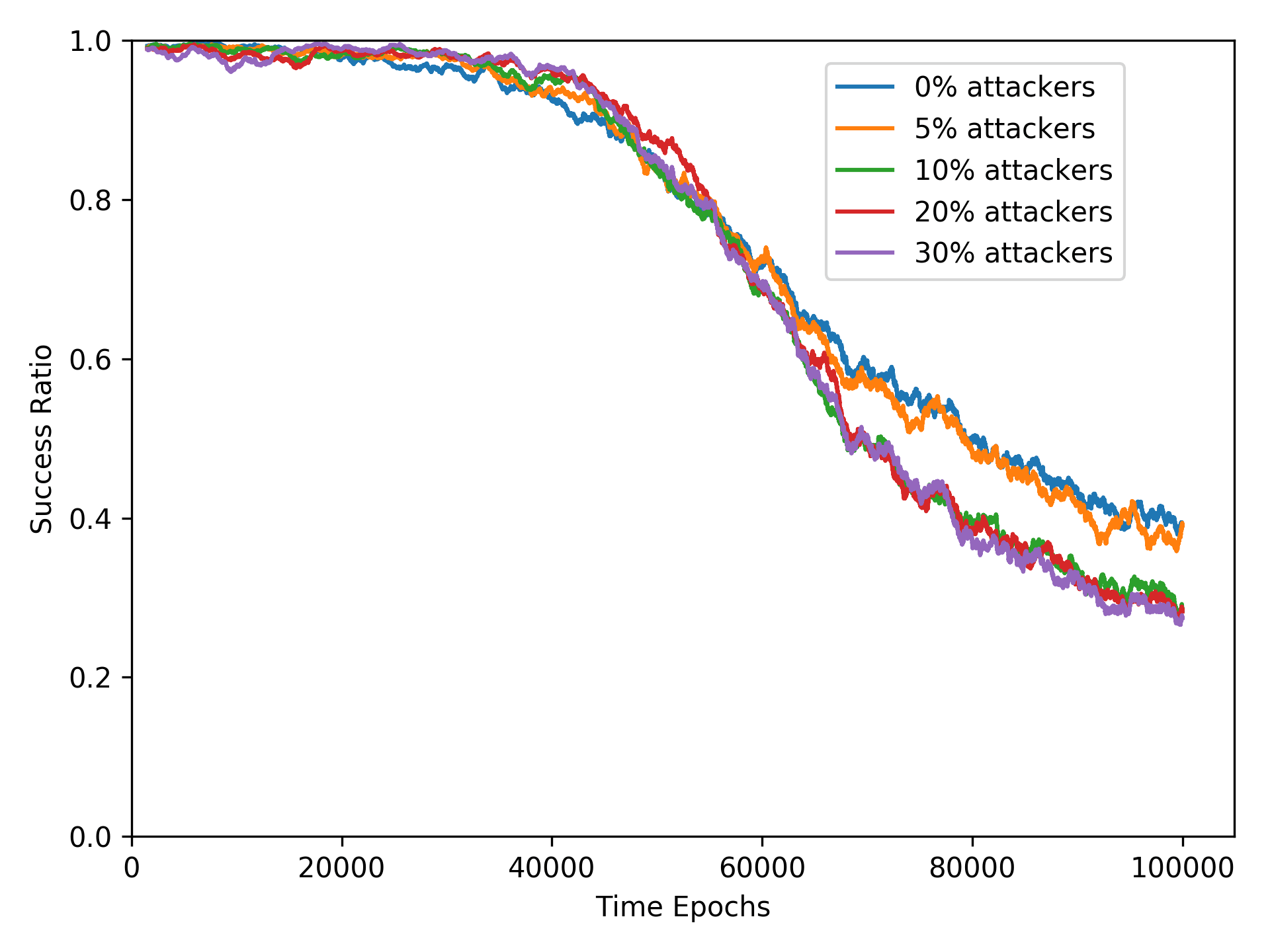}{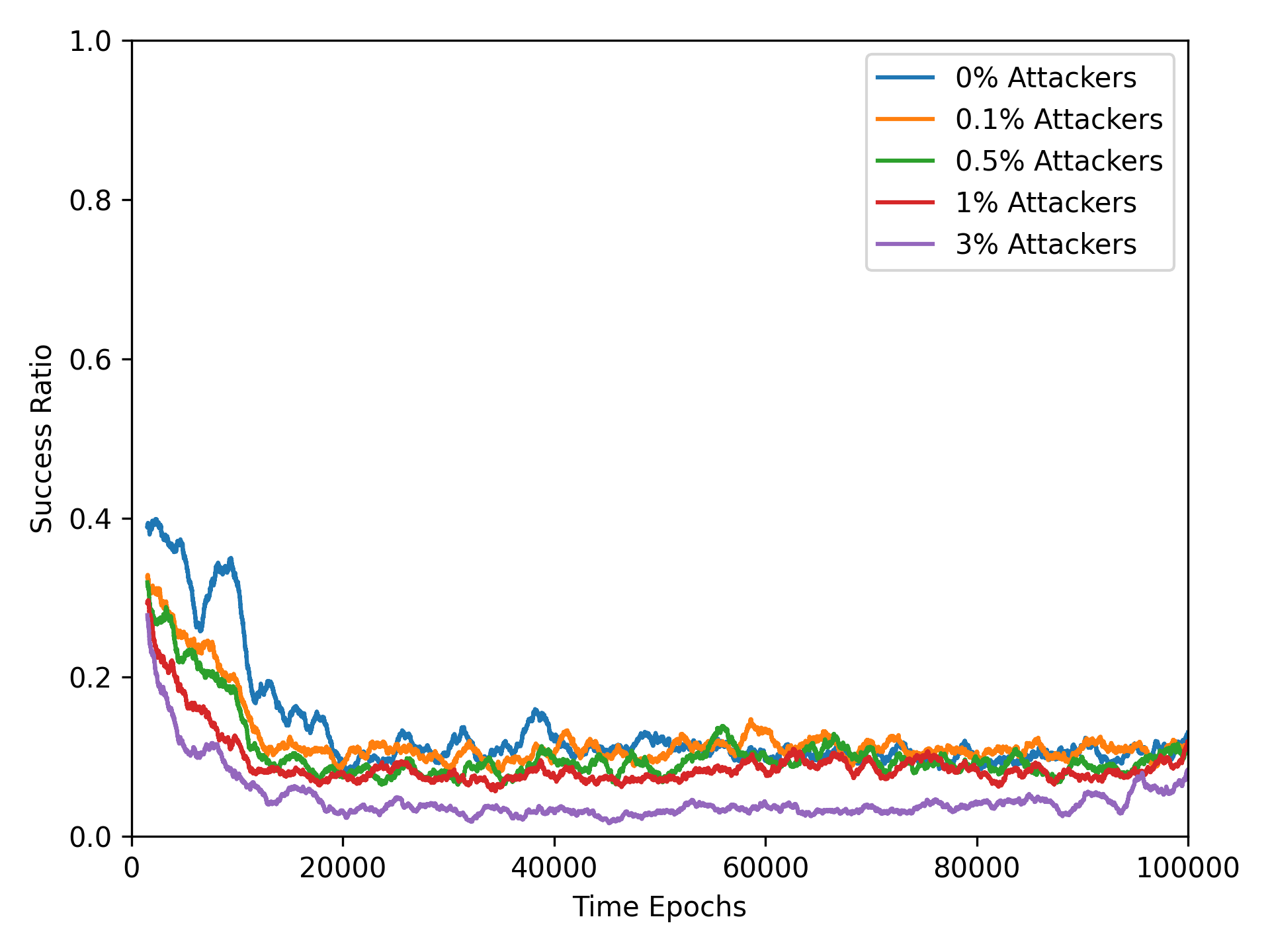}{Success
  ratios during a griefing by delaying attack. }

We present the results of our tree-based attacker in
Fig.~\ref{figure:id25-exp57-drop_all-success-ratio}. This attack is specific to
the spanning-tree structure of the \sm routing algorithm, and therefore cannot
be attempted against non-tree based algorithms such as \ff. The results show a
much stronger attack than when nodes are chosen randomly. We start to see a
noticeable drop in performance with only 0.5\% attackers, and with
 1\% attackers, the performance has dropped to only slightly above zero.
The difference in attack severity results from the
tree-based attack strategically selecting nodes with a high probability of being involved in many transactions.

Fig.~\ref{figure:id25-exp22-speedymurmurs-griefing_success-id25-exp22-maxflow-griefing_success-id25-exp51-griefing_success}
shows the results of our griefing by delaying attack. Against \ff
(Fig.~\ref{figure:id25-exp22-maxflow-griefing_success-success-ratio}), the
griefing by delaying attack shows no effectiveness until the later epochs, and
requires at least 10\% of the network to be attackers.
Fig.~\ref{figure:id25-exp22-speedymurmurs-griefing_success-success-ratio} shows
the results from performing this attack against \sm. Having 5\% attackers leads
to a slightly lower success ratio initially but then stabilizes at roughly the
same success ratio as a network without adversaries. A higher fraction of
attackers leads to a lower success ratio but follows the same pattern of an
initially high success ratio dropping and then stabilizing. At some point, the
effect per additional attacker seems to decrease. For instance, 30\%
attackers is only barely more effective than 20\%. This is not surprising,
because once such large swaths of the network are controlled by attackers, most
paths already contain one adversary so that an additional attacker does not
increase the number of affected paths.

The tree-based attack is more
effective than a random one, the random attacker requires 30\% of the nodes
to achieve the same impact as the tree-based attack with only 3\%.

\begin{figure} []
  \begin{subfigure}{\figwidth\columnwidth}
    \centering
    \includegraphics[width=\columnwidth]{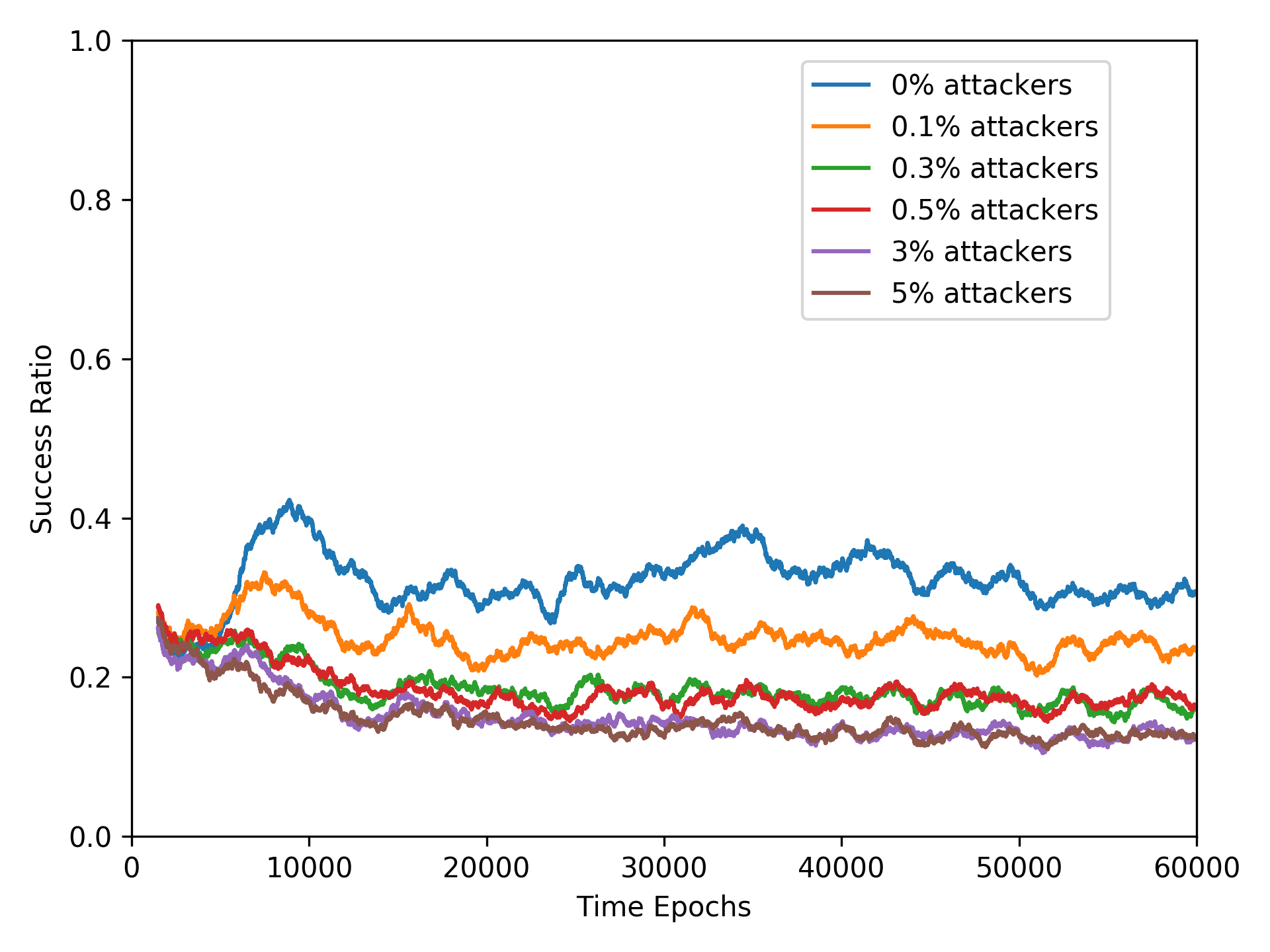}
    \caption{\sm}
    \label{figure:sm-ln}
  \end{subfigure}
  \begin{subfigure}{\figwidth\columnwidth}
    \centering
    \includegraphics[width=\columnwidth]{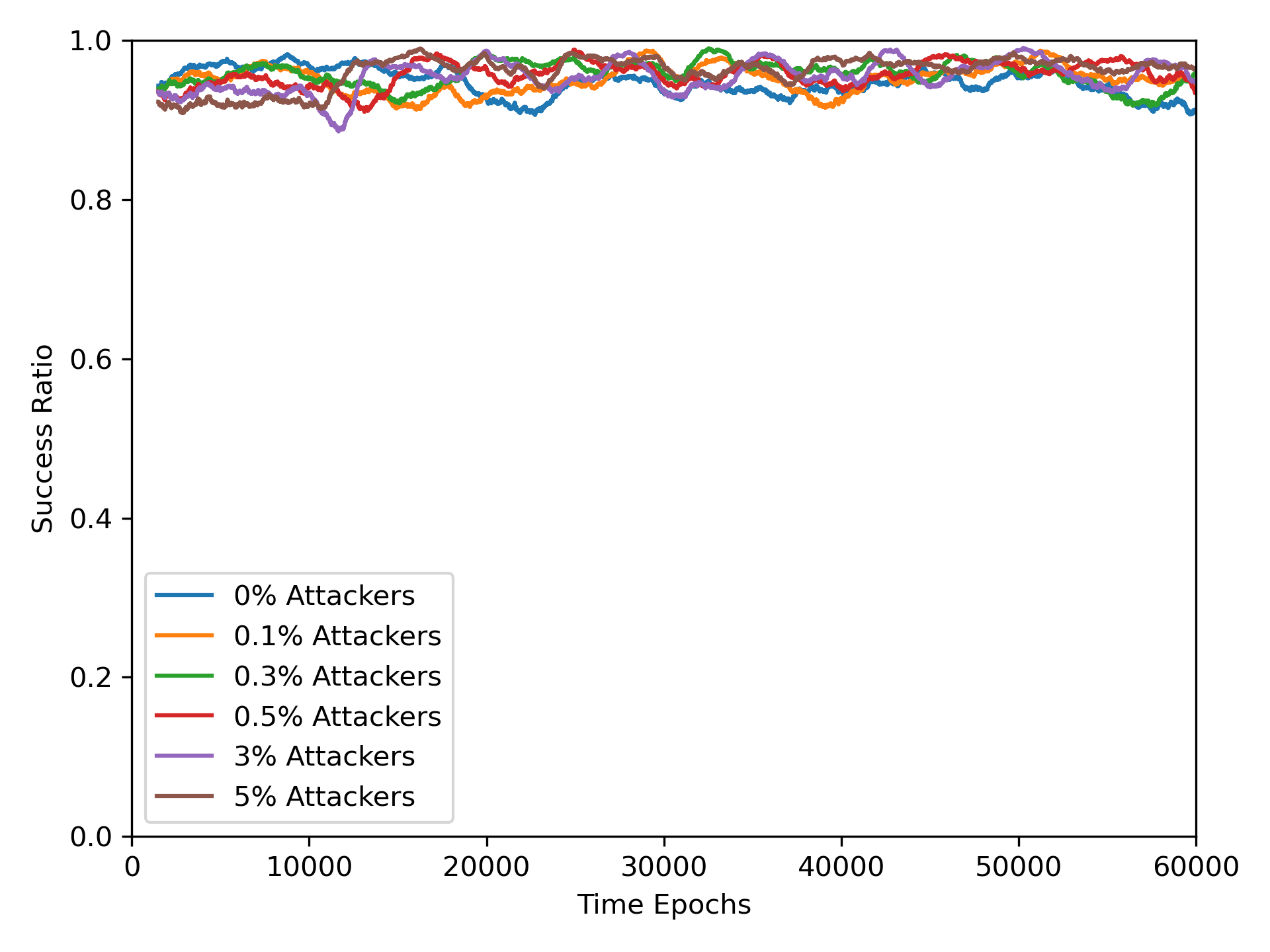}
    \caption{\ff}
    \label{figure:sm-ln}
  \end{subfigure}

  \caption{Griefing attack against the Lightning topology with generated
    transactions. Attackers selected randomly.}
  \label{figure:griefing-ln}
  \vspace{-0.5em}
\end{figure}

We also present our result for the dropping and griefing by delaying attacks on the Lightning
Network topology. Due to space constraints, we include only the griefing attack
with attackers selected randomly (Fig.~\ref{figure:griefing-ln}) --- however, in
all cases, the results were consistent with the attacks against synthetic
datasets.

We observe, in all scenarios, that dropping is a more effective attack than
griefing by delaying. However, network operators can more easily detect it
than the griefing attack. This is because in a griefing attack, a
transaction may fail due to an attacker delaying a transaction on a partially
overlapping path, whereas with dropping, the transaction must be directly routed
through the attacker's node.

\subsection{Centrality-based Attacker Nodes Selection}
If the attacker has the ability to strategically choose which nodes to corrupt,
it can use the available topology information to corrupt the nodes with the
highest betweenness centrality. We next evaluate our graph-oriented attack for
both \sm and \ff
(Fig.~\ref{figure:id25-exp23-speedymurmurs-drop_all-id25-exp23-maxflow-drop_all}).
Against \sm, this attack reaches maximum efficacy with only 0.1\% attackers. As
the success ratio is brought down to zero by so few attackers, adding more
attackers does not offer any additional benefit. We see a similar increase in
efficacy for \ff in Fig.~\ref{figure:id25-exp23-maxflow-drop_all-success-ratio}.
With just 0.1\% attackers, the success ratio drops more than 70\% in the early
epochs. More attackers offer diminishing returns, but still bring down the
success ratio to almost zero with only 5\% attackers.

\plotexpsCentrality{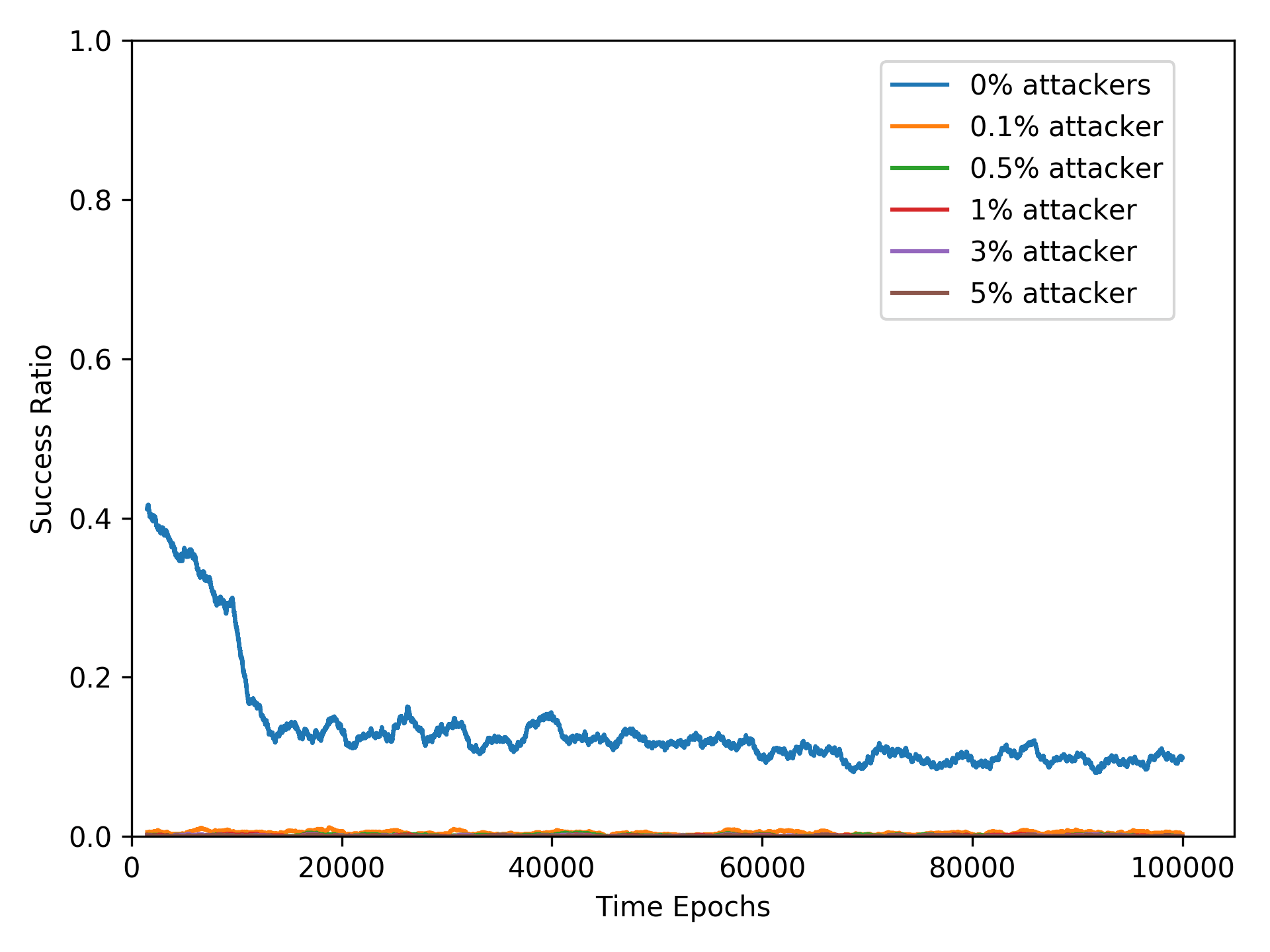}{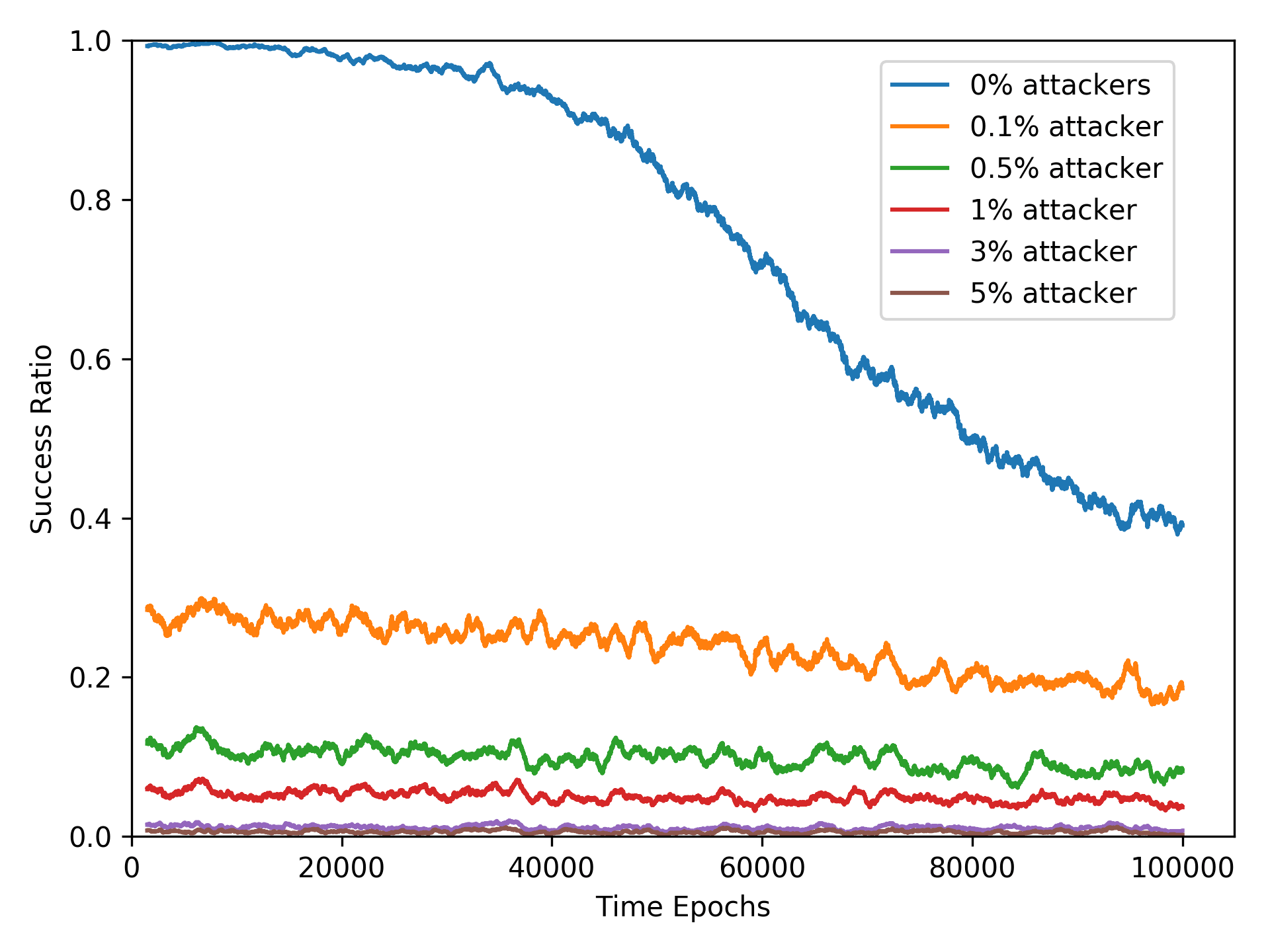}{Success
  ratios during a dropping attack when selecting attackers by betweenness
  centrality. }

Fig.~\ref{figure:id25-exp23-speedymurmurs-griefing_success-id25-exp23-maxflow-griefing_success}
displays the effects of griefing by delaying against both \sm and \ff when
selecting attackers by highest betweenness centrality. For \sm
(Fig.~\ref{figure:id25-exp23-speedymurmurs-griefing_success-success-ratio}), 0.1\%
attackers are sufficient to reduce the success ratio to almost zero. Given that
the success ratio is already so low, additional attackers do not considerably
decrease it further. The effects on \ff are less pronounced
(Fig.~\ref{figure:id25-exp23-maxflow-griefing_success-success-ratio}), as it
can more easily route around failures.

\plotexpsCentrality{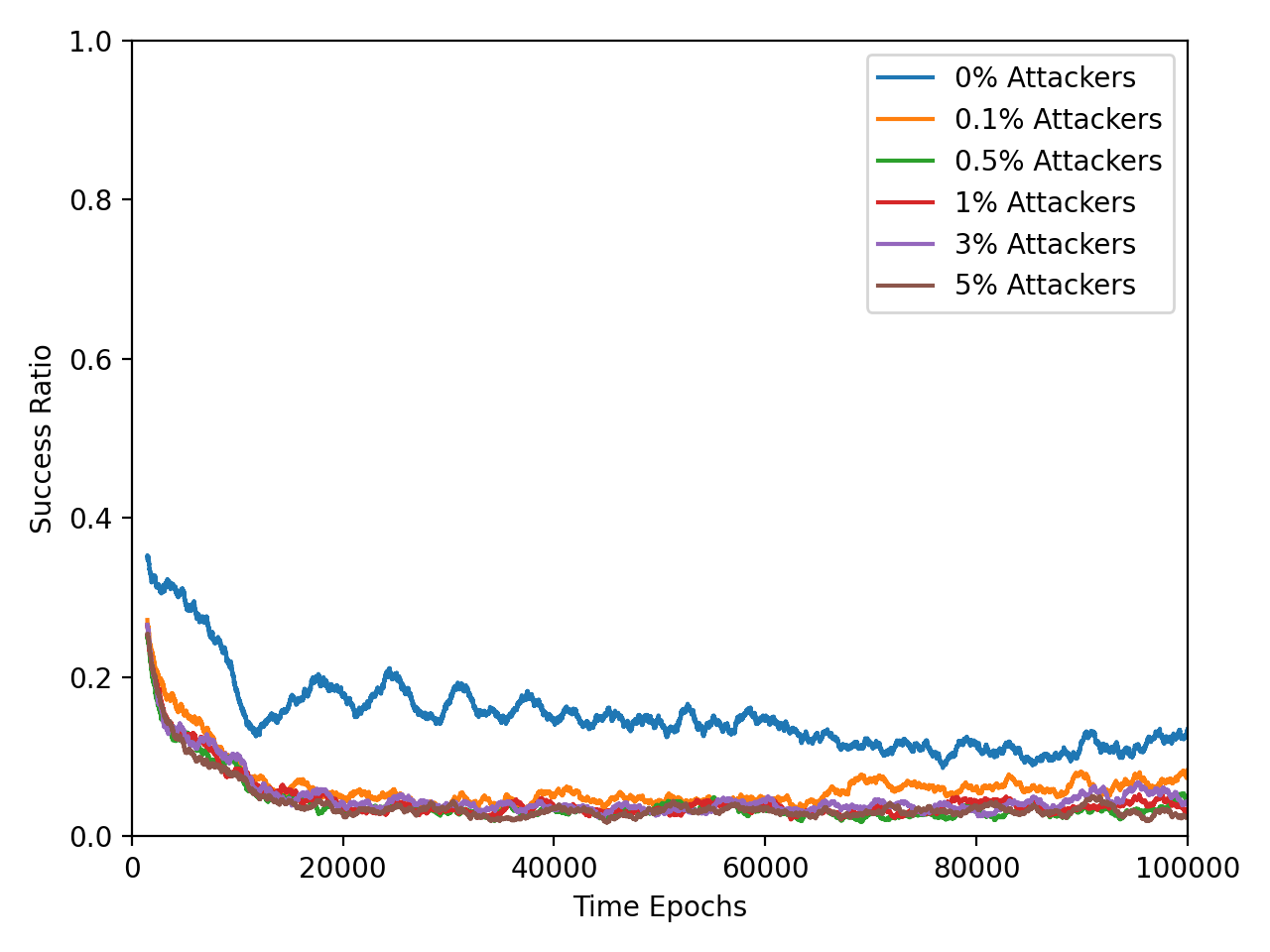}{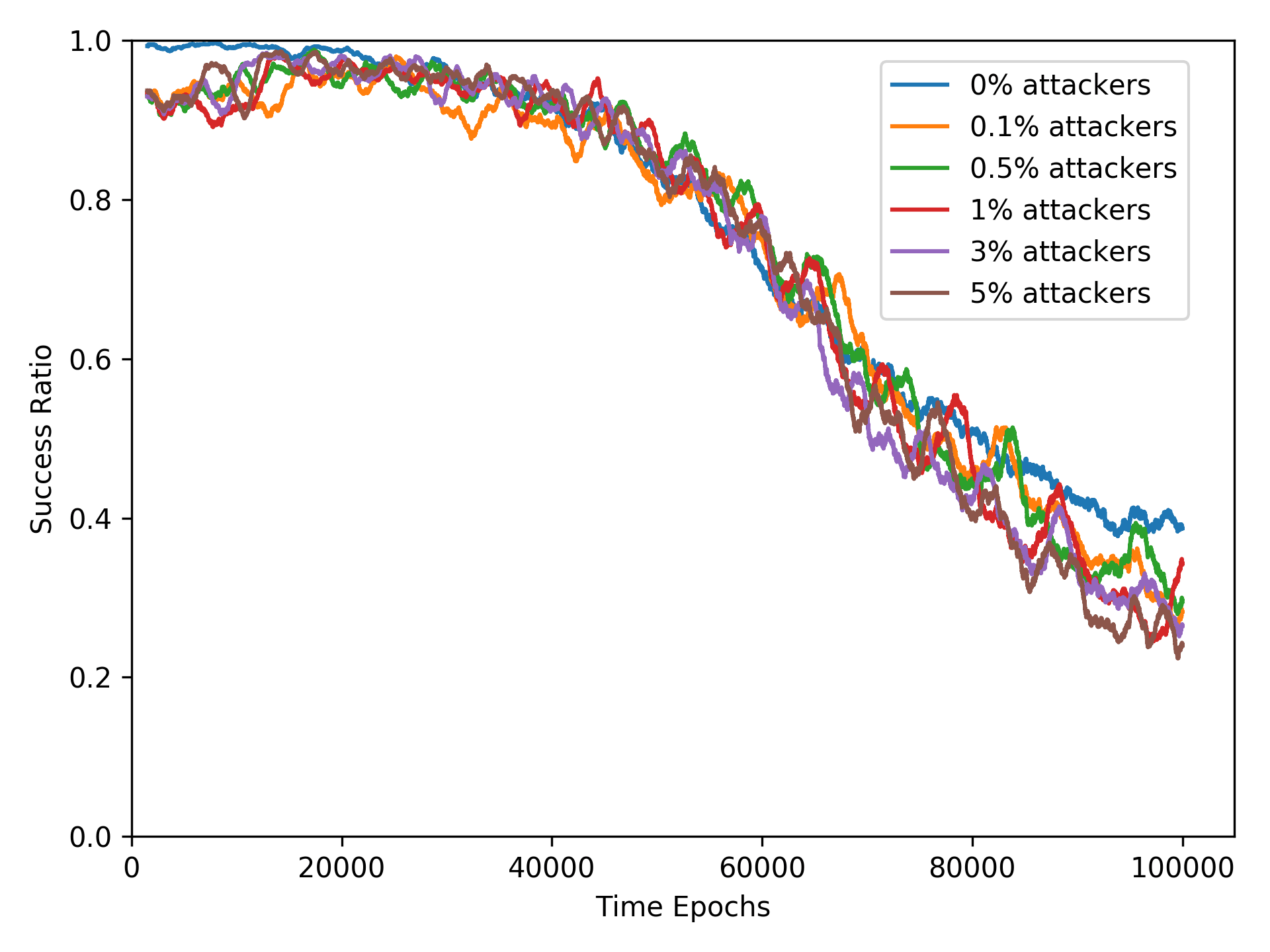}{Success
  ratios during a griefing by delaying attack when selecting attackers by
  betweenness centrality. }

The effectiveness of the centrality-based attacks indicates conclusively that a
node's position in the topology does impact attacker power. As \sm
is a tree-based algorithm, it may seem surprising that the tree-based attacker
is less effective than the tree-agnostic centrality-based attacker. However,
 the most important factor in whether an attacker
will be effective is how many transactions pass through it. This in turn is
affected by the \src and \dest locations within the topology. In our datasets,
we used a Poisson distribution for assigning the \src and \dest of each
transaction, which means that many of the transaction {\src}s and {\dest}s will
be concentrated in relatively few nodes. An attacker near one of those would be
much more likely to impact many payments. A transaction from a \src or to a
\dest farther from the root may be able take shortcuts before reaching the
attackers that are closer to the root (see Section~\ref{sec:routingalgo} for details
about \sm routing), which means that an attacker close to the root will not
necessarily be involved in more transactions than other nodes. A
centrality-based attacker is not as affected by the distribution of payment
{\src}s and {\dest}s, because the high number of paths passing central nodes includes paths with short cuts as well as those without.

\plotexpsComparison{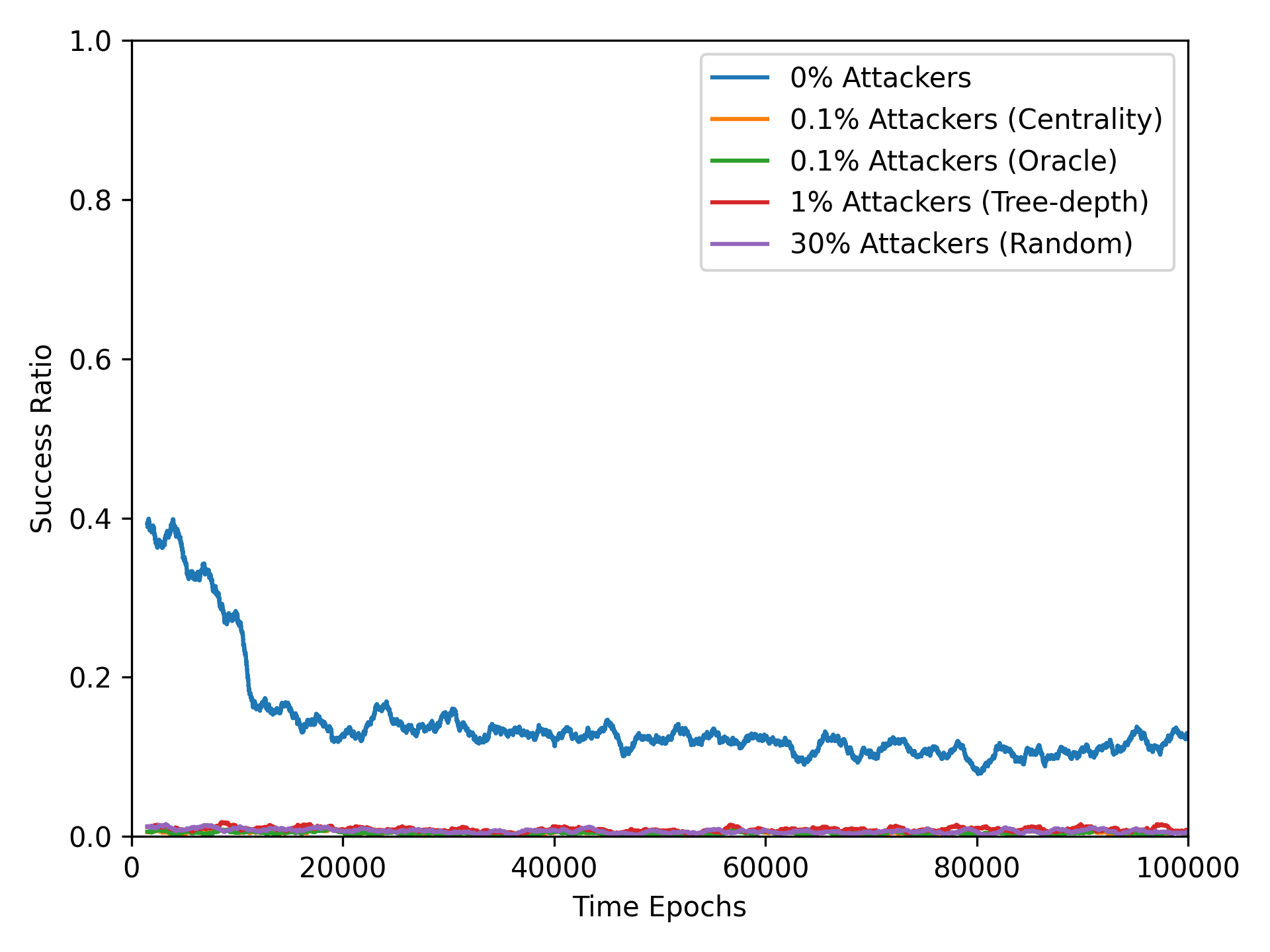}{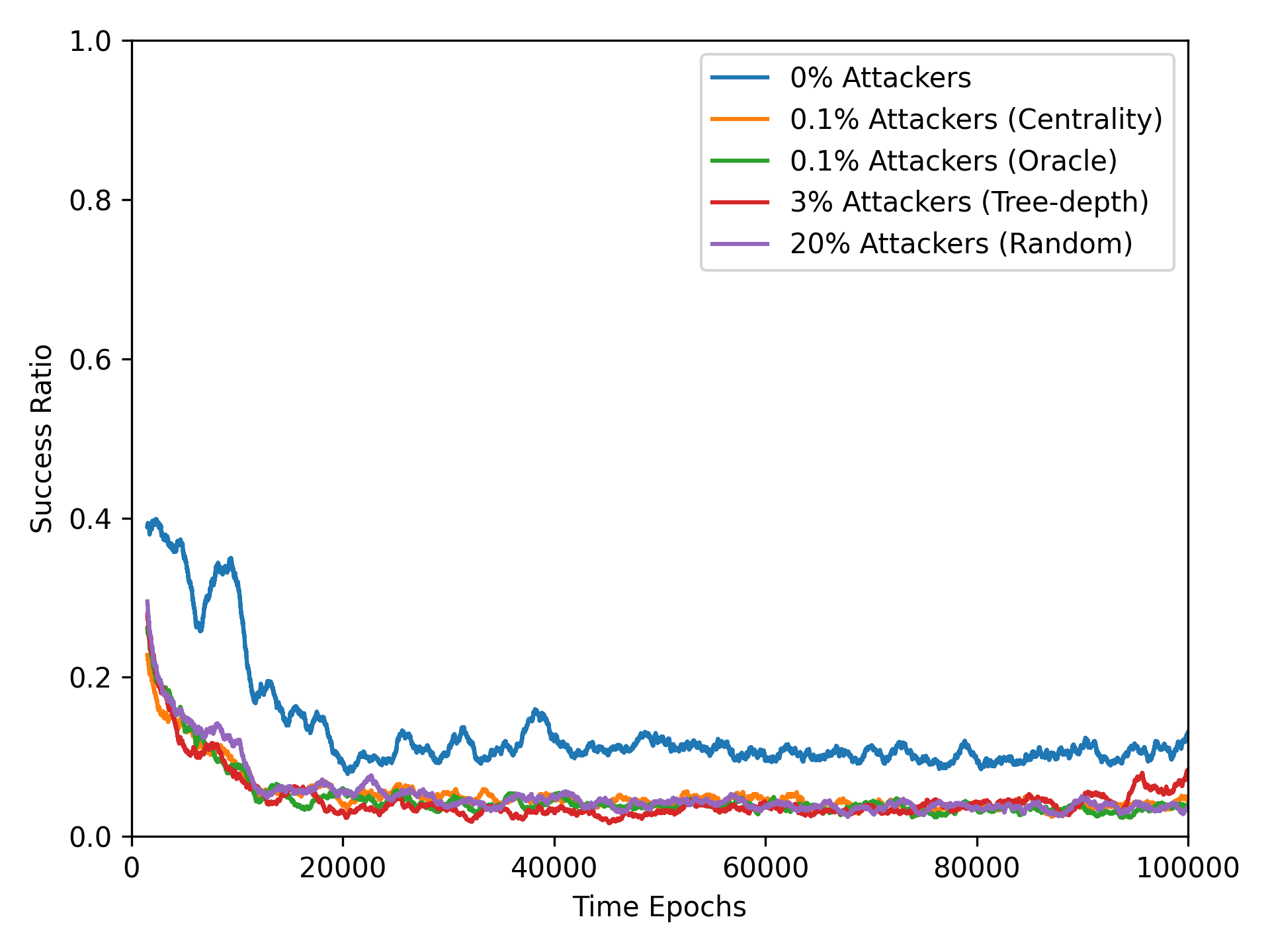}{Comparison
  of how many malicious nodes are needed to reach the maximum attack efficacy
  for dropping and griefing by delaying attacks.}

Fig.~\ref{figure:compare-selection-methods-dropping-drop_all-compare-selection-methods-griefing-griefing_success}
shows a comparison of the number of malicious nodes needed to make an attack
effective. With enough attackers, all three selection methods (random,
centrality, and tree-depth) achieve optimal efficacy, however, the
centrality-based attack is able to do so with nearly as few malicious nodes as
the oracle (0.1\%), whereas the tree-depth attack requires 3\% and the random
attacker requires 20\%.

% !TEX root = creditnetworks.tex

\subsection{Attack Cost}
There are two ways to insert corrupt participants into a payment channel
network: i) corrupting existing nodes or ii) inserting new malicious nodes into
the network. Both come at a monetary cost and probably relate to a party's
centrality, meaning that achieving a more central position is more expensive.
\newtext{We speculate this is the case because a node with more valuable channels
  may employ stringent security measures. This assumption only affects our cost
  analysis, and has no bearing on the attack.}

Intuitively, corrupting influential participants, \eg, through malware, should
be more costly if nodes invest in better protection, however, to the best of our
knowledge, there is no study on the costs of protections employed by blockchain
users. Hence, we focus on the scenario when the attacker
integrates nodes into the network by establishing channels.

\paragraph{Centrality-based Attacks} As in other
approaches~\cite{ersoy2020profit}, we assume that honest nodes are willing to
accept requests to open channels, which offer more payment opportunities and
potential income in the form of fees, but only if the party initiating the
channel opening provides all necessary funds. Thus, for each channel, the
attacker has to provide the blockchain fee for a channel opening $f^t_{open}$
with $t$ indicating that the fee depends on the time of the opening. In
addition, they have to supply the channel capacity $cap$, which has to be high
enough that parties choose this channel. In Lightning, the average and median
channel capacity are 0.03 and 0.05 Bitcoin, which is more than 500 and 900 US
dollars as of Dec 6, 2020~\footnote{\url{https://1ml.com/statistics}}. So, $cap$
should be on the order of hundreds of dollars. In comparison, $f^t_{open}$
varies from about half a dollar to more than 10 dollars. Hence, the cost to
initialize $K$ channels of the same capacity at time $t$ is
\begin{equation}
  cost_{BET}(K) = K\cdot (f^t_{open} + cap).
\end{equation}
When closing the channel, the adversary regains the funds locked in his
direction, \ie, funds they can still spend. However, for a node to forward
a payment to the attacker, funds in the other direction are needed, meaning that
the attacker has to make payments to achieve a suitable channel balance,
optimally with most of the funds available to its partners. As a consequence,
the attacker is unlikely to receive a considerable amount of the invested funds
back.
For our experiments, the total number of channels of the adversary is $K=1376$  for 0.1\% attackers, \ie, the attacker amount in Fig.~\ref{figure:compare-selection-methods-griefing-griefing_success-success-ratio}, meaning the attack cost is close to 1 million US dollars when inserting new nodes.

\paragraph{Tree-based Attacks} In a tree-based attack, the attacker can
establish a channel with another node and immediately tear it down upon learning
that the node is not a suitable partner due to its position in the tree. Let $X$
be the random variable indicating the number of connections attempts needed to
connect to a root node. In addition, let $f^t_{close}$ be the fee for closing
the channel. Hence, if a node aims to establish $T$ connections in total with
one connection to a root node, the expected cost of the attack are
\begin{equation}
  cost_{Tree}(T) = T\cdot (f^t_{open} + cap) + \mathbb{E}(X)(f^t_{open}+f^t_{close}).
\end{equation}
We observe an average of $T=3307.7$ for 3\% attackers in our experiments. The random variable $X$ depends on the node's knowledge of the topology. Therefore, to achieve the same effect as the centrality-based attack, the cost for the tree-based attack is higher.

% !TEX root = creditnetworks.tex

\subsection{Mitigation}
We focus on the centrality-based attack as it was the most effective attack.
Intuitively, the absence of nodes with a high centrality should reduce the attack severity. We first check if this intuition holds true and then name incentives for nodes to form networks with less central nodes.

\expSuccPLCreditCumulative{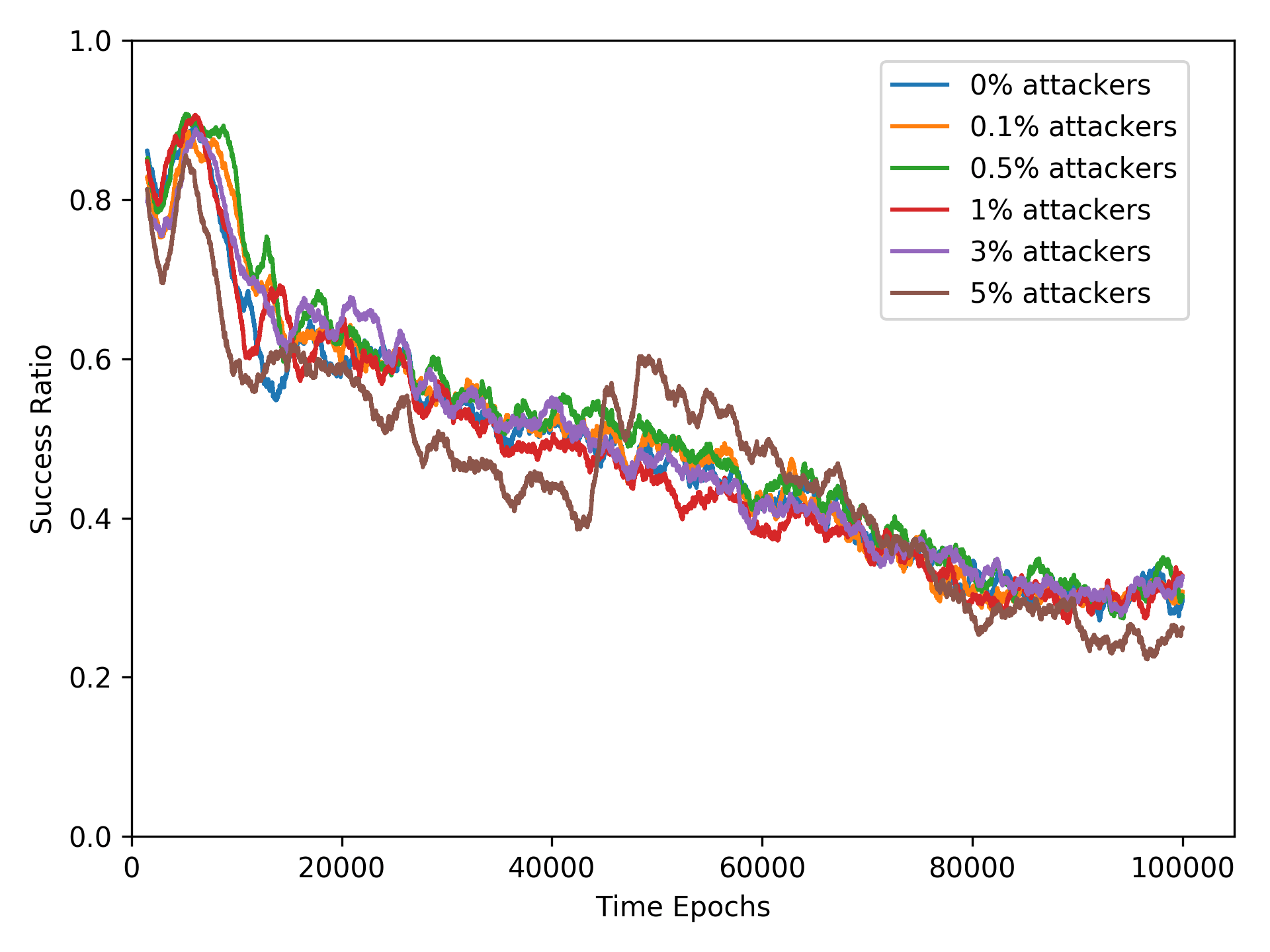}{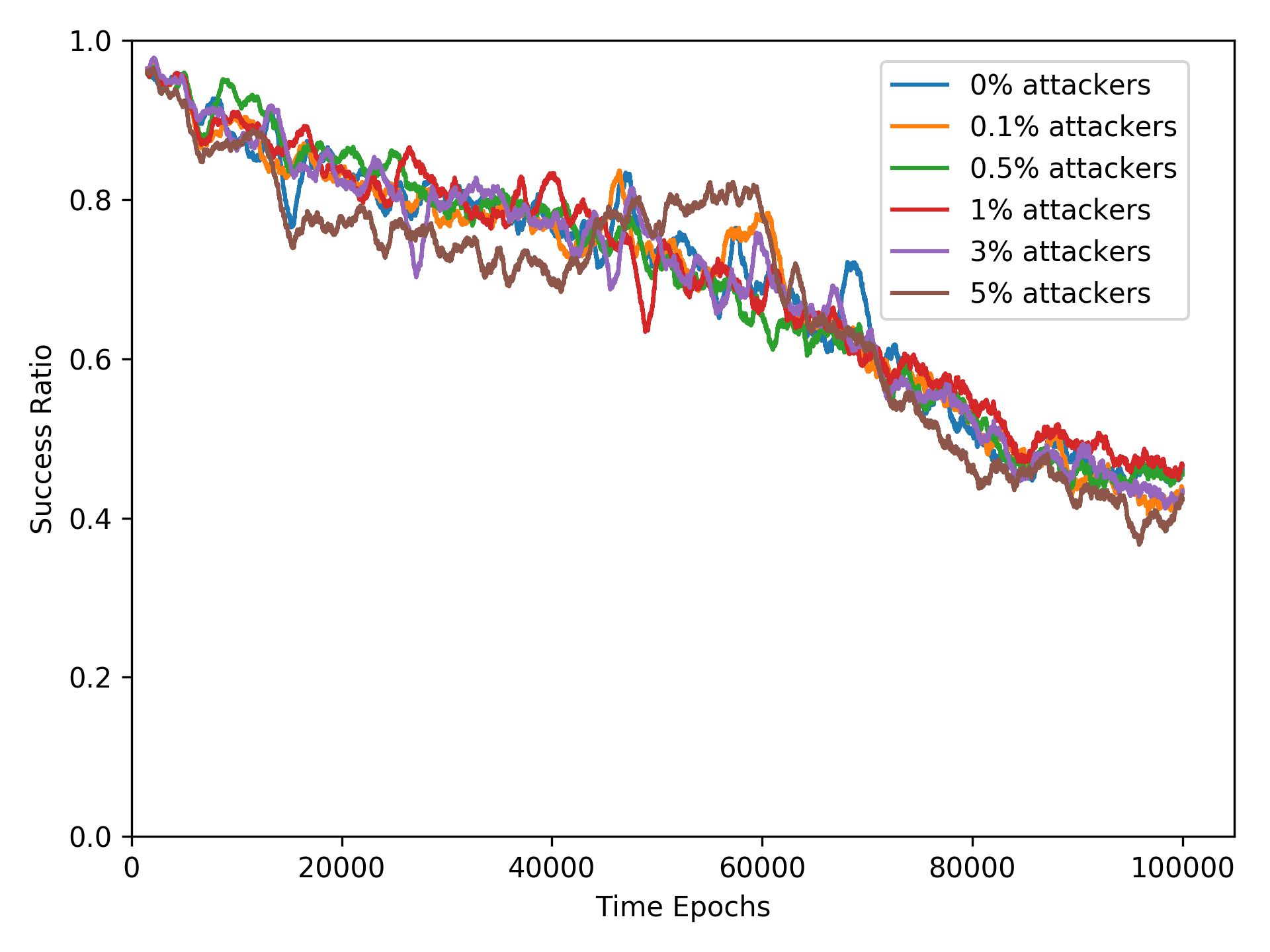}{Mitigation against griefing by delaying; attackers selected by betweenness centrality.}{exp30-griefing}

We generate a
Newman-Watts-Strogatz (NWS) network~\cite{newman1999renormalization}, a connected small-world
graph where all nodes have a very similar degree.
The parameters for the NWS algorithm are $p$, the probability of a node adding a connection
to a node that is not one of its neighbors, and $k$, the number of close neighbors
to connect to. We choose a value of 0.01 for $p$ and 20 for $k$.
For assigning initial balances, we used the algorithm from Section~\ref{sec:datasets}, with the addendum that if
a channel had a balance below some minimum value, then we increased it to that minimum value to avoid having too many channels with no capacities. Channels of capacity 0 are unlikely due to the fact that opening a channel costs a fee.

As expected, the effect of our attacks are considerably lower in a small-world network with attackers selected by highest betweenness centrality.
For the griefing attack, the attack has almost no impact for both \sm and \ff,
as can be seen in Figure~\ref{figure:exp30-griefing}.
When dropping, there is necessarily an effect as at least the dropped payments fail. However, the fraction of payments routed via the nodes with the highest betweenness centrality are tiny in number in comparison to the scale-free network.
Hence, the effect of dropping only becomes visible when a higher number of nodes is corrupted, as can be seen in Figure~\ref{figure:exp30-dropping}  for 5\%.

Thus, incentivizing a more homogeneous topology is a suitable mitigation.
A simple incentive is to warn participants not to form channels with strangers
that are willing to pay the complete fees as such behavior often precedes an
attack. Indeed, there are blockchain congestion attacks that can lead to
monetary loss due to delayed disputes that have a similar attack
setup~\cite{harris2020flood}, further emphasizing the need for preventing an
attacker from establishing channels to random nodes.

\expSuccPLCreditCumulative{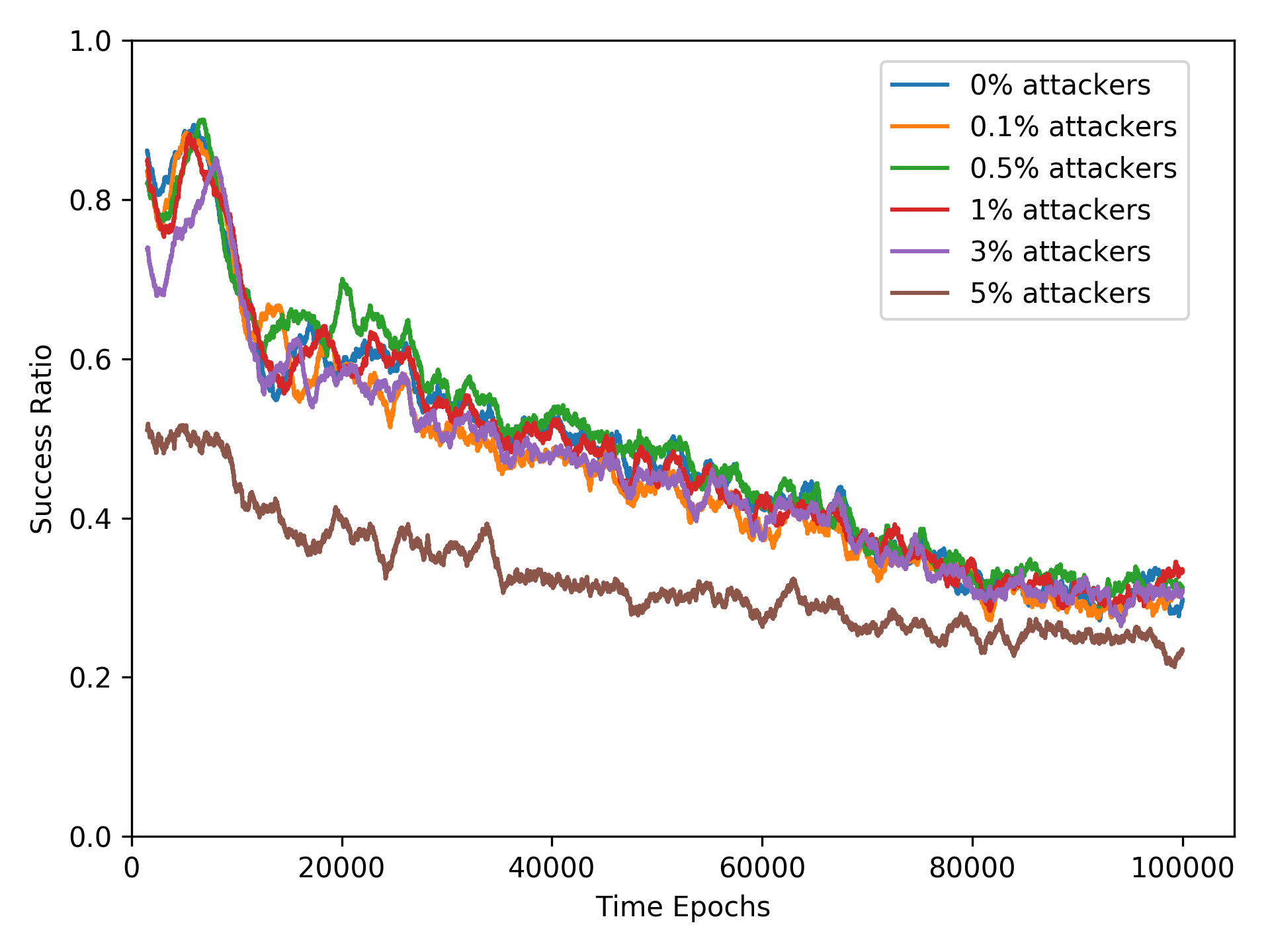}{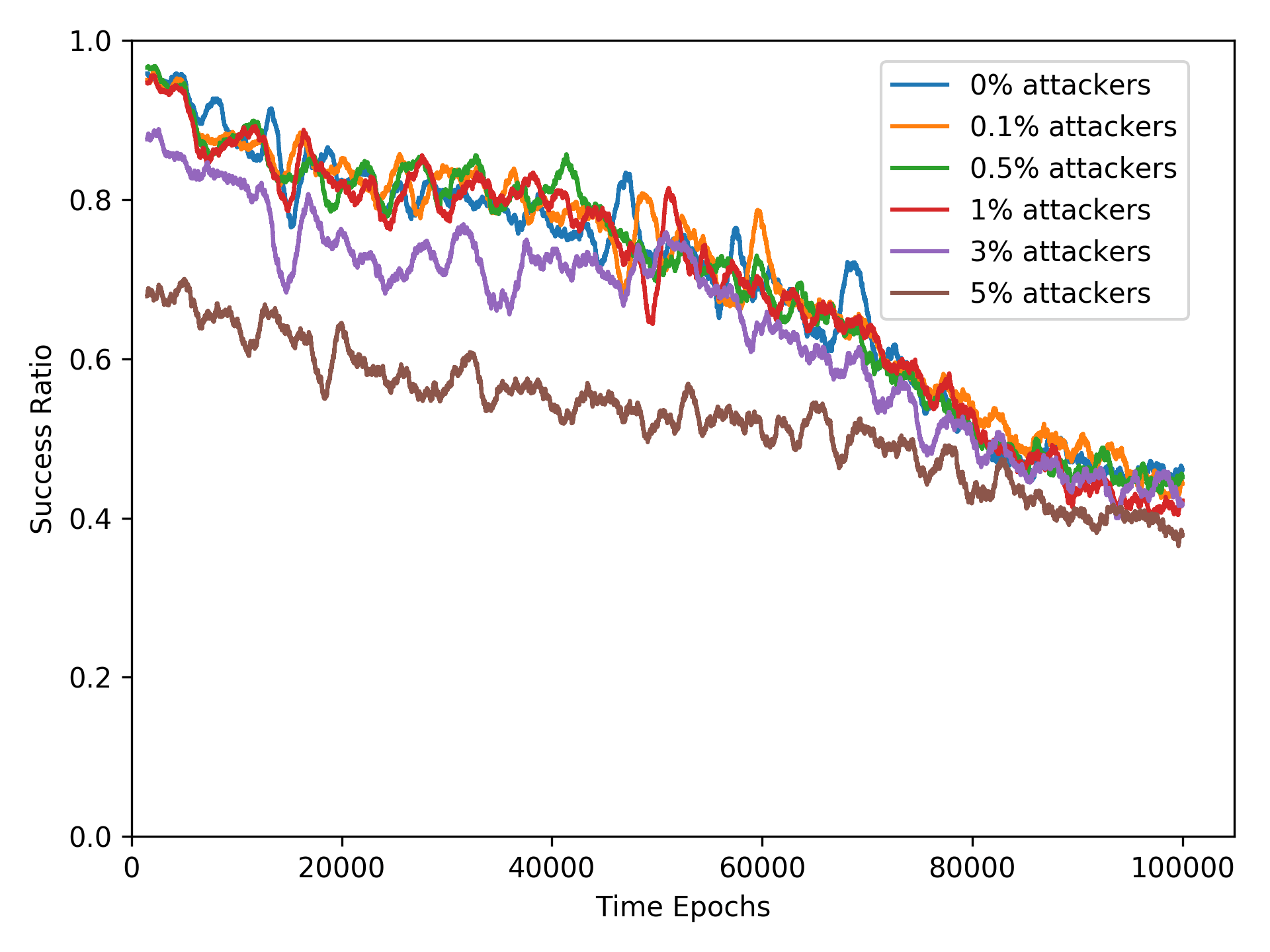}{Mitigation against dropping; attackers selected by betweenness centrality.}{exp30-dropping}

% !TEX root = paper.tex

\section{Related Work}

We refer to an in-depth survey for a detailed review of payment channel networks~\cite{gudgeon2019sok} and focus on the work related to attacks and defense mechanisms.
%Furthermore, we analyzed the differences between our data sets and previously used data sets.
%The existing work on attacks on availability all focus on source routing and specifically its use in Lightning.

{\bf Payment Griefing and Channel Exhaustion.}
Rohrer \etal introduce various attacks on Lightning, namely denial-of-service, channel exhaustion, payment griefing and node isolation, as well as combinations of these~\cite{rohrer2019discharged}.
\sandb{Their}\sandbrm{The} attacks \sandbrm{presented in [7]}leverage the same broad ideas as ours but focus on fundamentally different routing algorithms, datasets, and threat model. They focus on the attacker initiating payments rather than affecting payments which they are on the path of. \sandbrm{Initiating a large number of payments is likely to cause suspicion and hence our attacks are more stealthy.} Our attacker model is more realistic as well. Rohrer {\etal}'s \emph{node isolation} assumes that a node will not take action if its channels are nearing depletion. In fact,  a node can rebalance its channels using either the blockchain or a circular payment.
Perez \etal design and analyze a more effective payment griefing attack~\cite{perez2019lockdown}.
However, the attack is not applicable for local routing algorithms.

The effects of payment griefing can be mitigated by replacing the \newtext{LN's
  collateral-locking protocol, Hashed-Timelock Contracts (HTLC)}. Lightning's current protocol might lock a payment for time $\mathcal{O}(l \Delta)$ if the path length is $l$ and an upper bound for initiating a dispute is $\Delta$.
Several approaches could reduce this maximal timeout ~\cite{miller2019sprites,egger2019atomic,jourenko2021payment}.
As all approaches still require timeouts in the order of minutes, our griefing by delaying attack remains applicable.

{\bf Dropping and congestion.}
Dropping attacks have been evaluated in the context of Lightning, finding that Lightning's hub and spoke topology as well as the predictable routing algorithm results in a high rate of failed payments if attackers are integrated into the network strategically~\cite{tochner2019hijacking}.
Protections against dropping are  randomization~\cite{tochner2019hijacking} and redundancy~\cite{bagaria2019boomerang}.
%Evaluating our attacks in the presence of payments with redundancy is an interesting venue for future work.

An alternative attack is network congestion~\cite{mizrahi2020congestion}.
In addition to the above attacks on Lightning, there are attacks on its privacy, indicating that channel balances and payment relations can be revealed~\cite{herrera2019difficulty,tikhomirov2020probing,kappos2020empirical}. Suddenly closing many channels can lead to congestion on the underlying blockchain and hence loss of funds due to disputes not being raised~\cite{harris2020flood}.

\section{Conclusion}

We examined the effectiveness of attacks against
payment channel networks that use local routing to complete
transactions. 
The performance of such routing algorithms degrades gracefully when the number of randomly placed attackers increases. However, they are very vulnerable if the attacker controls 
nodes of a high centrality and hence controls the majority of paths. 
We propose to incentivize payment channel networks with homogeneous node degrees. 

\section*{Acknowledgments}
\sandb{The authors would like to thank Matthew Jagielski for his help designing
  the balance allocation algorithm, and Jaison Titus for many insightful
  discussions.}

\bibliographystyle{IEEEtran}
\bibliography{creditnetworks}

\appendix
% !TEX root = creditnetworks.tex

\vspace{-5pt}
\section{Generating Synthetic Datasets}
\label{sec:datasets}
The effectiveness of PCN routing algorithms, and attacks
against them, is highly dependent on the topology, channel balances, transactions pairs,
and transaction-value distributions.
We discuss limitations of existing datasets and then describe our approach.

\vspace{-5pt}
\subsection{Limitations of Existing Datasets}
Due to the lack of available data sets for PCNs, early
approaches~\cite{speedymurmurs,silentwhispers} relied on data from Ripple's
credit network~\cite{ripple} for topology, capacities, and transactions. This
dataset was collected by crawling the Ripple network, taking note of which
accounts are funded, (\ie, can make payments to other accounts) and removing inconsistencies.
%that may have arisen from ignoring certain Ripple-specific features, like  currency exchange by intermediaries.
Early approaches not based on Ripple do not consider transaction and capacity distributions at all~\cite{prihodko2016flare}.

%
%Much of the work on evaluating payment channel networks relies on a publicly
%available dataset from measured from Ripple \cite{speedymurmurs,sivaraman2020high}.
%This dataset was collected by crawling the Ripple network, taking note of which
%accounts are funded (\ie can make payments to other accounts), and removing inconsistencies
%that may have arisen from ignoring certain Ripple-specific features, like exchanging
%currencies.
%%% \cnr{add one sentence describing the dataset}.

\begin{figure} []
  \centering
  \begin{subfigure}[t]{\figwidth\columnwidth}
    \includegraphics[width=\columnwidth]{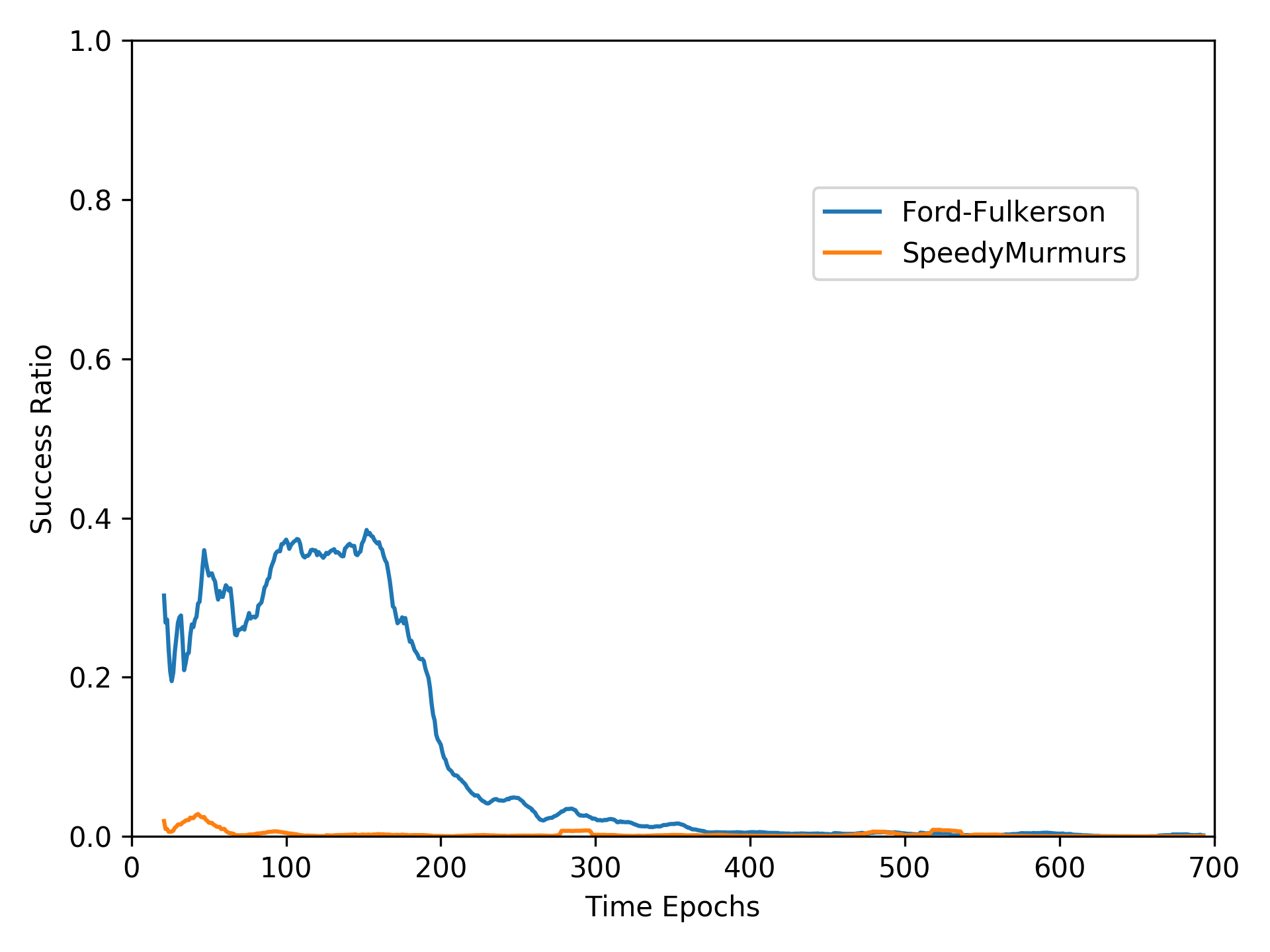}
    \caption{Success ratio.}
    \label{fig:baseline-sequential-compare-algos-id3_s}
  \end{subfigure}
  \begin{subfigure}[t]{\figwidth\columnwidth}
    \centering
    \includegraphics[width=\columnwidth]{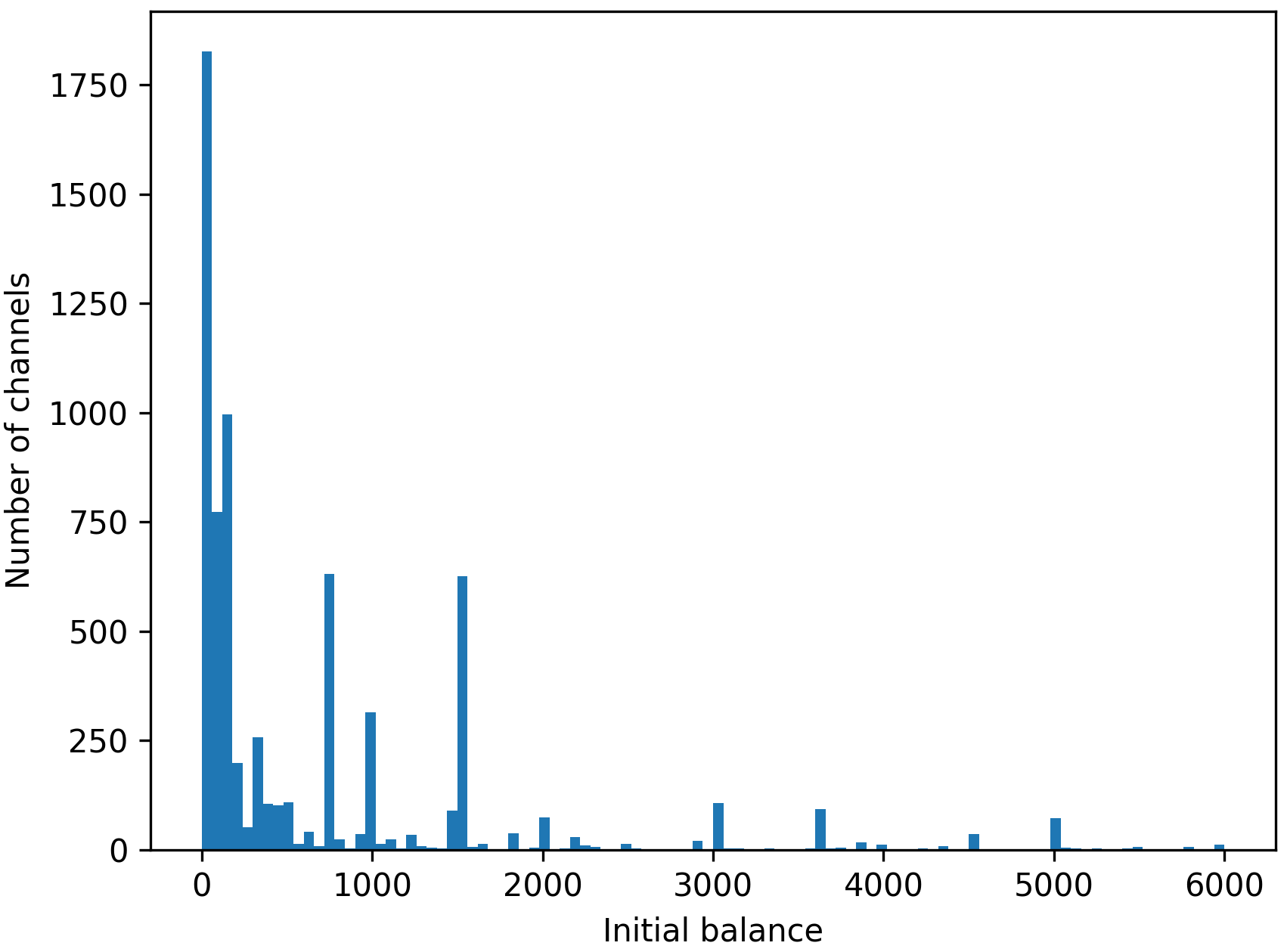}
    \caption{Initial balances}
    \label{figure:id3-topology_CREDIT_LINKS_s}
  \end{subfigure}

  \caption{Ripple dataset. On the left, performance of \ff and \sm with sequential transactions and no attack,
  on the right initial balances.}
  \label{figure:ripple}
\end{figure}

Unfortunately, the Ripple dataset has a very small ratio of transactions that
are successful as seen in Figure~\ref{fig:baseline-sequential-compare-algos-id3_s},
where we show \sm and \ff with sequential transactions.
%\ie, when completing each transaction before starting the next one.
Performance for concurrent transactions would be even worse.
Previous work addresses the issue by excluding transactions that were not successful for \ff, leading to a very limited transaction set~\cite{speedymurmurs}.
As shown in Figure~\ref{figure:id3-topology_CREDIT_LINKS_s}, the majority of initial channel balances are zero or very close to zero. While zero balances make sense for a credit network, where credit corresponds to trust and channel establishment does not come at a fee, it is unrealistic in the context of payment networks such as Lightning. Opening a channel requires paying the fees for one blockchain transaction, which is the same regardless of the amount locked in the channel. Thus, channels with little or no funds are unattractive as they do not provide the opportunity to make payments and hence users are unlikely to invest the fee for opening them.
%We present further details about the Ripple dataset in Figure~\ref{figure:ds3}.

%\dsfig{3}{Ripple Dataset Characteristics.}

After Lightning data became available, many studies relied on the Lightning topology and capacities for their evaluation~\cite{sivaraman2020high,rohrer2019discharged,perez2019lockdown,tochner2019hijacking,
mizrahi2020congestion}, although some used synthetic random graphs and scale-free topologies either as their main dataset~\cite{bagaria2019boomerang} or in addition to Lightning data~\cite{sivaraman2020high}.

The Lightning data set does not contain transactions. A dataset for transactions
requires both \src-\dest pairs and values. Prior works have chosen \src-\dest
pairs uniformly at
random~\cite{bagaria2019boomerang,tochner2019hijacking,rohrer2019discharged}.
Only the evaluation of the Spider routing algorithm uses an exponential
distribution~\cite{sivaraman2020high}, indicating that few {\src}s and {\dest}s
are very active whereas the majority of nodes participate only occasionally.
Transaction values have been modeled on real-world data unrelated PCNs,
\eg, Ripple transactions in~\cite{rohrer2019discharged,bagaria2019boomerang} and
credit card transactions in~\cite{sivaraman2020high,cordi2017simulating}. Using
data from the Raiden Network~\cite{raidennetwork} is a non-starter as it only
has 37 unique accounts~\cite{RaidenExplorer}.

In summary, there are no real-world datasets available for payment channel
networks capturing all the information about transactions and with realistic
initial balance sets. As a result, it has become common in the literature to use
synthetic graphs in lieu of the smaller LN graphs in addition to the synthetic
transactions~\cite{bagaria2019boomerang,sivaraman2020high}.

%, thus we need to generate synthetic datasets.
%Our data sets are different in that they allow us to vary the expected success ratio and hence see the difference between a network whose capacity is clearly sufficient to handle transactions and one that has a limited capacity.

\subsection{Our Approach}
A payment channel system is defined by the network graph, $G$, created by the channels
between participants in the system,
the transactions performed between these participants, $T$,
the set of balances available on each channel, $B$,
and the routing algorithm, $R$, used to find payment paths; we write $PC = <G, T, B, R>$.
$B$ is initialized with a set of initial balances $B_0$.
Our high-level approach to generate datasets is as follows:

(1) Channel network: We first generate the channel network, modeled
as a graph  $G = <N, C, B>$, where $N$ is the set of participants, $C$
the set  of channels, and $B$ is the set of balances on each channel.
Note that $B$ is undefined at this point, we will assign an initial balance set
$B_0$ later. We use scale-free graphs as they were
shown to be representative for PCNs~\cite{rohrer2019discharged} and
we write $G = SF(n, c)$,  where $SF$ is the  scale-free graph generation algorithm,
$n$ is the total number of participants and $c$ is a connectivity parameter
that models how many channels a party forms when added to
the graph.

(2) Transaction set: We then select a transaction set.
A transaction is defined by a pair of \src
and \dest and the value of the transaction. We separate the selection of the transaction
pair from the selection of the value of the transactions and write
$T = GT(N, nt, v_{fix}, D_v, D_n)$, where $GT$ is our procedure of generating a transaction set and
$T = \{t_i, t_i= <s_i,r_i,v_i>\}$ is the resulting transaction set with
$s_i$ and $r_i$ as the \src and \dest, respectively, for transaction $t_i$ of value $v_i$.
The parameters of $GT$ are the set of participants, $N$, the number of transactions, $nt$,
sampled from distribution $D_v$, $v_{fix}$ is an additional parameter determining $D_v$ (\eg, minimum, maximum, average, etc.), and
$D_n$ is the distribution used to sample pairs of parties.

(3) Initial balance set: Finally, we generate the set of initial balances for
each channel. Our approach starts with a balance set of channels with capacity 0, a
given set of transactions, and a percentage, $tc$, of transactions for which we want to have the guarantee that they can be successfully completed.
We iterate over the set of transactions, and, with probability $tc$, execute the routing algorithm $R$ to find one or several channels in the graph between \src and \dest. For each channel along the paths returned by $R$, we
add the transaction value $v$ to the balance. In this manner, we know that there is a possibility for the transaction to be successful.
Formally, we write $B_0 = CIB(T, G, R, tc, m_c, p_c)$ and $B = \{b_i, b_i = (s_i, r_i) \mapsto v_i\}$,
where $CIB$ is our algorithm to compute the initial balance set,
$T$ is  transaction set, $G$ is the channel network,
$R$ is the routing algorithm and $tc$ is the percentage of completed transactions.
In addition, we might reduce the final channel balance by a factor, $m_c$, to generate a network with less liquidity. $p_c$ is the probability of applying such a multiplier to a channel, \eg, $p_c=1$ indicates the multiplication is applied to all channels.
We also write $P_i,G_{i+1}=R(t_i,G_{i})$, where routing a transaction $t_i$ using
algorithm $R$ on network $G_{i}$ results in finding a path $P_i = \{c, c = <e_1,e_2,b>\}$, and a network with
updated balances, $G_{i+1}$. For each channel in $P_i$, $e_1$ and $e_2$ are the endpoints of the channel, and
$b$ is the value being routed through that channel.

%Below we describe in more details the generation of the transaction set and of the initial
%balance set.

\subsection{Generating Datasets}

\subsubsection{Generating Transaction Sets}
\label{sec:txsets}
Each transaction is a tuple of the transaction value, the \src, and the \dest, with the latter two forming the transaction pair. We treat the selection of transaction values and transaction pairs independently.

\textit{Transaction value.} We consider five distributions: constant, Pareto,
exponential, normal, and Poisson. For the experiments presented below, we only
show results using Pareto distributions as it has been shown to most reliably
represent consumer spending patterns~\cite{sandersPARETOPRINCIPLEITS1987}.

\textit{Transaction pair.} We sample the \src and \dest independently according
to a distribution $D_n$. For this, we first randomize the order of the nodes and
then map each of them to the index they have in this random order. We then
sample the node based on its index according to the distribution. In particular,
we focus on the Poisson distribution as existing literature has shown it to be
the best model for consumer transaction
rates~\cite{faderCountingYourCustomers2005}.

\subsubsection{Generating Initial Balance Set}
\label{sec:cib}
%\begin{assumption}
%  The initial balance of a channel represents its expected use.
%  \label{ass:initisgood}
%\end{assumption}
%Assumption~\ref{ass:initisgood} has several implications.

The initial balance of a channel represents its expected use. Channels with
higher balances are expected to have a higher volume of transacted funds and
vice versa for lower balances. This is a result of the opportunity cost to
escrowing funds in payment channels. A rational actor will want to participate
in as many transactions as possible to collect fees. Also, larger channels will
permit transactions that may have been too large for smaller channels to route,
thus participating in more transactions. The ratio between the balance from $A \rightarrow
B$ to the balance from $B \rightarrow A$ represents the ratio of transacted funds from
$A \rightarrow B$ to the transacted funds from $B \rightarrow A$.

We assign balances to channels as follows. We consider a PCN, which is some
graph $G_0=(N=\lbrace u\rbrace, C=\lbrace (u, v)\rbrace)$. A transaction in a PCN can occur between a
\src, $s$, a \dest, $r$, and with a value, $v$---the full transaction $t_i$ is
written as $(s_i,r_i,v_i)$. A routing algorithm, $R$, is run on a graph, $G$,
and with a transaction $t_i$; it returns a route, $P_i$, which is the set of all
modified channels, and also $G_{i+1}$, the graph after all channels are
modified. We write this as $P_i,G_{i+1}=R(t_i,G_i)$. The routing algorithm, $R$,
may return multiple paths and may also be randomized. If a route with sufficient
balances cannot be found, \ie, there is not a high enough funds on the channels
connecting $s_i$ to $r_i$, $R$ returns $\bot$ and $G_i$ ($G_i$ is not modified).
The modified graph, $G_{i+1}$, is equivalent to, for each hop $(a, b, m) \in P$,
either subtracting $m$ from the channel $(a,b)$'s funds in $G$ or adding $m$
to the channel $(b,a)$'s funds.

\textit{Reducing Graph Weights.}
Generating the perfect channel balances is useful for testing the system, but to simulate certain concurrency scenarios, we need some of the channels to have insufficient credit to complete \emph{all} of the transactions in $T$. To achieve this, we simply take all edge weights in the graph, and scale them by some multiplier, \ie, $E' = \{(u,v,k\cdot w)\}_{(u,v,w) \in E}$, where $k$ is a constant scale factor.

We also experimented with the option of only applying the factor $k$ probabilistically, \ie, applying it only with probability $p_c$. A complementary approach for reducing the capacity of the network is to select transactions for the generation of the initial balance generation randomly, with probability $tc$.

\subsubsection{Generated Datasets}
We used the implementation of the Barab\'{a}si-Albert
algorithm~\cite{barabasi1999emergence} for the scale-free graph generation from
the Python module networkx~\cite{networkx}. The datasets were generated
according to the methodologies discussed above and are shown in
Table~\ref{tab:datasetsFull}. In deciding the number of nodes to model, we
settled on a value of 10k as it was larger size than the current number of
active LN nodes but still small enough that it would reflect the network in the
near future.
%% \stef{Number of nodes now seems to be more than 10k but
  %% number of actve nodes still slightly below: \url{https://1ml.com/statistics}}
%Each simulation is run on a unique transaction set sampled in accordance with section~\ref{sec:txsets}.
%The transaction set used in the full-knowledge weight assignment algorithm  is a transaction set uniquely generated for each graph instance unless stated otherwise.
%The algorithm used for finding the path between \src and \dest to increase the initial balance was always a simple shortest path algorithm.
While we experimented with all datasets in the table,
we present our results for \dsSF.

\begin{table}[]
  \scalebox{0.7}{
    \centering
    \tabcolsep=0.015\linewidth
        \begin{tabular}{@{}l|l|l|l@{}}
      \toprule
      Name      & Topology             & Transaction Set                                   & Initial Balance Set \\ \midrule
      0         & SF(100k, 5)          &    GT(100k, 1m, 1, Pareto, Pareto)                & CIB(T, G, shortest path, 100\%, 1, 1) \\
      6         & SF(100k, 5)          &    GT(100k, 1m, 1, Pareto, Pareto)                & CIB(T, G, shortest path, 100\%, 0.5, 1) \\
      7         & SF(100k, 2)          &    GT(100k, 1m, 1, Pareto, Pareto)                & CIB(T, G, shortest path, 100\%) \\
      8         & SF(10k, 2)           &    GT(10k, 1m, 1, Pareto, Pareto)                 & CIB(T, G, shortest path, 100\%, 0.5, 1)\\
      10        & SF(10k, 2)           &    GT(10k, 1m, 1, Pareto, Pareto)                 & CIB(T, G, shortest path, 100\%, 1, 1) \\
      12        & SF(10k, 2)           &    GT(10k, 1m, 1, Pareto, Constant)               & CIB(T, G, shortest path, 100\%, 1, 1) \\
      13        & SF(10k, 2)           &    GT(10k, 1m, 1, Poisson, Poisson)               & CIB(T, G, shortest path, 100\%, 1, 1) \\
      14        & SF(10k, 2)           &    GT(10k, 1m, 1, Exp, Exp)       & CIB(T, G, shortest path, 100\%, 1, 1) \\
      15        & SF(10k, 2)           &    GT(10k, 1m, 1, Poisson, Poisson)               & CIB(T, G, shortest path, 100\%, 0.5, 0.5)\\
      16        & SF(10k, 2)           &    GT(10k, 1m, 1, Normal, Normal)                 & CIB(T, G, shortest path, 100\%, 1, 1) \\
      17        & SF(10k, 2)           &    GT(10k, 1m, 1, Normal, Normal)                 & CIB(T, G, shortest path, 100\%, 0.5, 0.5)\\
      18        & SF(10k, 2)           &    GT(10k, 1m, 1, Normal, Normal)                 & CIB(T, G, shortest path, 100\%, 0.5, 0.5)\\
      19        & SF(10k, 2)           &    GT(10k, 1m, 1, Normal, Normal)                 & CIB(T, G, shortest path, 80\%, 1, 1)  \\
      20        & SF(10k, 2)           &    GT(10k, 1m, 1, Poisson, Pareto)                & CIB(T, G, shortest path, 100\%, 1, 1) \\
      21        & SF(10k, 2)           &    GT(10k, 1m, 30, Poisson, Normal)               & CIB(T, G, shortest path, 100\%, 1, 1) \\
      22        & SF(10k, 2)           &    GT(10k, 1m, 1, Poisson, Constant)              & CIB(T, G, shortest path, 100\%, 1, 1) \\
      23        & SF(10k, 2)           &    GT(10k, 1m, 1, Poisson, Pareto)                & CIB(T, G, shortest path, 100\%, 0.5, 0.5)\\
      24        & SF(10k, 2)           &    GT(10k, 1m, 1, Poisson, Pareto)                & CIB(T, G, shortest path, 100\%, 0.5, 1)\\
      \hl{Scale-Free}        & \hl{SF(10k, 2)}           &    \hl{GT(10k, 100k, 1, Poisson, Pareto)}                & \hl{CIB(T, G, shortest path, 100\%, 0.5, 0.5)}\\
      %% 26        & SF(100k, 5)          &    GT(100k, 1m, 1, Poisson, Pareto)                & CIB(T, G, shortest path, 100\%, 0.5, 0.5)\\
      %% 27        & ER(10k, N/A, 0.001)  &    GT(10k, 100k, 1, Poisson, Pareto)              & CIB(T, G, shortest path, 100\%, 0.5, 1)\\
      %% 28        & WS(10k, 10, 0.001)   &    GT(10k, 100k, 1, Poisson, Pareto)              & CIB(T, G, shortest path, 100\%, 0.5, 1)\\
      \hl{Small-World}        & \hl{NWS(10k, 20, 0.01)}    &    \hl{GT(10k, 100k, 1, Poisson, Pareto)}              & \hl{CIB(T, G, shortest path, 100\%, 0.5, 1, 100)}\\
      \hl{Lightning}        & \hl{Measured}    &    \hl{GT(2337, 100k, 1, Poisson, Pareto)}              & \hl{CIB(T, G, shortest path, 100\%, 1, 1, 100)}\\
      \bottomrule
    \end{tabular}
  }
\caption{Description of datasets we generated and experimented with. Unless
  otherwise stated, our simulations use \dsSF. \sandb{The datasets analyzed in
    this paper are highlighted.}}
\label{tab:datasetsFull}
\end{table}

\end{document}